\documentclass[11pt]{article}
\usepackage{amsfonts}
\usepackage{amssymb}
\usepackage{amsmath}
\usepackage{graphicx}

\newcommand{\ket}[1]{|#1\rangle}

\newcommand{\bra}[1]{\langle #1|}

\newcommand{\bracket}[2]{\langle #1|#2\rangle}

\newcommand{\op}[1]{\operatorname{#1}}
\def \qed {\hfill \rule{0.2cm}{0.2cm}\vspace{3mm}}
\newenvironment{mylist}[1]
{\begin{list}{}{\setlength{\leftmargin}{#1}
\setlength{\rightmargin}{0.0cm}\setlength{\labelsep}{1.3mm}
\setlength{\labelwidth}{0.8cm}\setlength{\itemsep}{0.2cm}}}
{\end{list}}

\newtheorem{theorem}{Theorem}
\newtheorem{lemma}[theorem]{Lemma}
\newtheorem{cor}[theorem]{Corollary}
\newtheorem{prop}[theorem]{Proposition}

\newtheorem{defn}{Definition}

\begin{document}

\begin{titlepage}

\vspace*{1.5cm}

\begin{center}
{\Large\bf Quantum statistical zero-knowledge}\\[8mm]
{\large
John Watrous\\
Department of Computer Science\\
University of Calgary\\
Calgary, Alberta, Canada\\
jwatrous@cpsc.ucalgary.ca}\\[4mm]
February 19, 2002
\end{center}

\vspace{2mm}

\begin{abstract}
In this paper we propose a definition for (honest verifier) quantum statistical
zero-knowledge interactive proof systems and study the resulting complexity
class, which we denote {QSZK}.
We prove several facts regarding this class:

\begin{itemize}
\item
The following natural problem is a complete promise problem for QSZK:
given instructions for preparing two mixed quantum states, are the states
close together or far apart in the trace norm metric?
By instructions for preparing a mixed quantum state we mean the description
of a quantum circuit that produces the mixed state on some specified
subset of its qubits, assuming all qubits are initially in the $\ket{0}$
state.
This problem is a quantum generalization of the complete promise problem of
Sahai and Vadhan \cite{SahaiV00} for (classical) statistical zero-knowledge.

\item
{QSZK} is closed under complement.

\item
{QSZK} $\subseteq$ {PSPACE}.
(At present it is not known if arbitrary quantum interactive proof
systems can be simulated in {PSPACE}, even for one-round proof systems.)

\item
Any honest verifier quantum statistical zero-knowledge proof system can be
parallelized to a two-message (i.e., one-round) honest verifier quantum
statistical zero-knowledge proof system.
(For arbitrary quantum interactive proof systems it is known how to parallelize
to three messages, but not two.)
Moreover, the one-round proof system can be taken to be such that the prover
sends only one qubit to the verifier in order to achieve completeness
and soundness error exponentially close to 0 and 1/2, respectively.
\end{itemize}

\vspace{1mm}

\noindent
These facts establish close connections between classical statistical
zero-knowledge and our definition for quantum statistical zero-knowledge, and
give some insight regarding the effect of this zero-knowledge restriction on
quantum interactive proof systems.

\end{abstract}

\end{titlepage}

%=============================================================================%

\section{Introduction}
\label{sec:introduction}

In recent years there has been an effort to better understand the potential
advantages offered by computational models based on the laws of quantum
physics as opposed to classical physics.
Examples of such advantages include: polynomial time quantum algorithms
for factoring, computing discrete logarithms, and several believed-to-be
intractable group-theoretic and number-theoretic problems \cite{CheungM01,
Hallgren02, IvanyosM+01, Kitaev95, Mosca99, Shor97, Watrous01};
information-theoretically secure quantum key-distribution\
\cite{BennettB84, ShorP00}; and exponentially more efficient quantum than
classical communication-complexity protocols \cite{Raz99}.
Equally important for understanding the power of quantum models are
upper bounds and impossibility proofs, such as the containment of
BQP (bounded error quantum polynomial time) in PP \cite{AdlemanD+97,
FortnowR99}, the impossibility of quantum bit commitment \cite{Mayers97},
and the existence of oracles relative to which quantum computers have
restricted power \cite{BennettB+97,FortnowR99}.

In this paper we consider whether quantum variants of zero-knowledge proof
systems offer any advantages over classical zero-knowledge proof systems.
Zero-knowledge proof systems were first defined by Goldwasser, Micali, and
Rackoff \cite{GoldwasserM+89} in 1985, are have since been studied
extensively in complexity theory and cryptography.
Familiarity with the basics of zero-knowledge proof systems is assumed in
this paper---readers not familiar with zero-knowledge proofs are referred to
Goldreich~\cite{Goldreich94,Goldreich99}.

Several notions of zero-knowledge have been studied in the literature, but we
will only consider {\em statistical} zero-knowledge in this paper.
Moreover, we will focus on {\em honest verifier} statistical zero-knowledge,
which means that it need only be possible for a polynomial-time simulator to
approximate the view of a verifier that follows the specified protocol (as
opposed to a verifier that may intentionally deviate from a given protocol
in order to gain knowledge).
In the classical case it has been proved that any honest verifier statistical
zero-knowledge proof system can be transformed into a statistical
zero-knowledge proof system against any verifier \cite{GoldreichS+98}.
The class of languages having statistical zero-knowledge proof systems
is denoted {SZK}; it is known that {SZK} is closed under complement
\cite{Okamoto00}, that {SZK} $\subseteq$ {AM} \cite{AielloH91,Fortnow89},
and that SZK has natural complete promise problems
\cite{GoldreichV99,SahaiV00}.
Several interesting problems such as Graph Isomorphism and Quadratic
Residuosity are known to be contained in {SZK} but are not known to
be in BPP \cite{GoldreichM+91,GoldwasserM+89}.
For further information on statistical zero-knowledge we refer the reader to
Okamoto \cite{Okamoto00}, Sahai and Vadhan~\cite{SahaiV00}, and
Vadhan~\cite{Vadhan99}.

To our knowledge, no formal definitions for {\em quantum} zero-knowledge proof
systems have previously appeared in the literature.
Despite this fact, the question of whether quantum models extend the class of
problems having zero-knowledge proofs has been addressed by several
researchers.
For instance, the applicability of bit-commitment to zero-knowledge proof
systems was one of the motivations behind investigating the possibility
of quantum bit commitment \cite{BrassardC+93}.
The primary reason for the lack of formal definitions seems to be that
difficulties arise when classical definitions for zero-knowledge are
translated to the quantum setting in the most straightforward ways.
More generally speaking, difficulties tend to arise in defining formal
notions of security for quantum cryptographic models (to say nothing
of proving security once a formal notion of security has been specified).
For a discussion of some of these difficulties, including issues specific to
quantum zero-knowledge, we refer the reader to van de Graaf \cite{Graaf97}.

We do not claim to resolve these difficulties in this paper, nor do we
propose a definition for quantum zero-knowledge that we feel to be satisfying
{\em from a cryptographic point of view}.
Rather, our goal is to study the complexity-theoretic aspects of a very
simple definition of quantum zero-knowledge based on the notion of an
honest verifier.
Our primary motives for considering this definition are as follows.
\begin{enumerate}
\item
Although we do not have satisfying definitions for quantum statistical
zero-knowledge when the honest verifier assumption is absent, it is obvious
that for any sensible definition that any quantum statistical zero-knowledge
proof system would necessarily satisfy our honest verifier definition.
Therefore, upper bounds on the power of honest verifier quantum zero-knowledge
proof systems also hold for the arbitrary verifier case.
(Our main results may be viewed as upper bound results.)

\item
We hope that by investigating simple notions of quantum zero-knowledge
we are taking steps toward the study and understanding of more
cryptographically meaningful formal definitions of quantum zero-knowledge
proof systems.

\item
We are interested in the effect of zero-knowledge-type restrictions on
the power of quantum interactive proof systems from a purely
complexity-theoretic point of view.
Indeed, we are able to prove some interesting facts about quantum
statistical zero-knowledge proof systems that are not known to hold for
arbitrary quantum interactive proofs, such as containment in PSPACE
and parallelizability to two messages.
\end{enumerate}

Our approach for studying a quantum variant of honest verifier statistical
zero-knowledge parallels the approach of Sahai and Vadhan \cite{SahaiV00} for
the classical case, which is based on the identification of a natural complete
promise problem for the class {SZK}.
We identify a complete promise problem for quantum statistical zero-knowledge
that generalizes Sahai and Vadhan's complete promise problem to the quantum
setting.
The problem, which we call the Quantum State Distinguishability problem, may
be informally stated as follows:
given instructions for preparing two mixed quantum states, are the states
close together or far apart in the trace norm metric?
The trace norm metric, which is discussed in more detail in the appendix, is
an extension of the statistical difference metric to quantum states, and gives
a natural way of measuring distances between quantum states.
By instructions for preparing a mixed quantum state we mean the description
of a quantum circuit that produces the mixed state on some specified
subset of its qubits, assuming all qubits are initially in the $\ket{0}$
state.
Naturally, the promise in this promise problem guarantees that the
two mixed states given are indeed either close together or far apart.

Several facts about quantum statistical zero-knowledge proof systems and the
resulting complexity class, which we denote QSZK, may be derived from
the completeness of this problem.
In particular, we prove that {QSZK} is closed under complement, that {QSZK}
$\subseteq$ {PSPACE} (which is not known to hold for quantum interactive proof
systems if the zero-knowledge condition is dropped, even in the case of
one-round proof systems), and that any honest verifier quantum statistical
zero-knowledge proof system can be parallelized to a one-round honest verifier
quantum statistical zero-knowledge proof system in which the prover sends only
one qubit to the verifier (in order to achieve completeness and soundness
error exponentially close to 0 and 1/2, respectively).

While our general approach follows the approach of Sahai and Vadhan, proofs
of several of the key technical facts differ significantly from the classical
case.
For instance, the proofs of completeness and closure under complement
rely heavily on properties of quantum states and thus have little resemblance
to the proofs for the classical analogues of these facts.

%-----------------------------------------------------------------------------%

\subsection*{Organization of the paper}

Section~\ref{sec:preliminaries} defines quantum interactive proof systems,
the quantum statistical zero-knowledge property, and the Quantum State
Distinguishability problem.
Section~\ref{sec:trace_distance} describes quantum zero-knowledge proof
systems for the Quantum State Distinguishability problem and its complement.
It is proved that the Quantum State Distinguishability problem is
complete for {QSZK} in Section~\ref{sec:completeness}, and various corollaries
of this fact as stated previously are stated more explicitly in this section.
We conclude with Section~\ref{sec:conclusion}, which mentions some
open problems regarding quantum zero-knowledge.
An overview of quantum circuits and some technical facts concerning the
quantum formalism are contained in an appendix that follows the main part of
the paper.

%=============================================================================%

\section{Preliminaries}
\label{sec:preliminaries}

%-----------------------------------------------------------------------------%

In this section we define quantum interactive proof systems, the quantum
statistical zero-knowledge property and the resulting class QSZK, and the
Quantum State Distinguishability problem which is shown to be complete for
QSZK in subsequent sections.
%Due to space limitations, we will not discuss quantum circuits or the quantum
%formalism in this section---an overview of these topics and references for
%further background information on quantum computing and quantum information
%may be found in the appendix.

%-----------------------------------------------------------------------------%

\subsection{Quantum interactive proofs}

Quantum interactive proofs were defined and studied in
\cite{KitaevW00,Watrous99-qip-focs}.
As in the classical case, a quantum interactive proof system consists of two
parties, a prover with unlimited computation power and a computationally
bounded verifier.
Quantum interactive proofs differ from classical interactive proofs in that
the prover and verifier may send and process quantum information.

Formally, a quantum verifier is a polynomial-time computable mapping $V$
where, for each input string $x$, $V(x)$ is interpreted as an encoding of a
$k(|x|)$-tuple $(V(x)_1,\ldots,V(x)_{k(|x|)})$ of quantum circuits.
These circuits represent the actions of the verifier at the different stages
of the protocol, and are assumed to obey the properties of polynomial-time
uniformly generated quantum circuits as discussed in the appendix.
The qubits upon which each circuit $V(x)_j$ acts are divided into two sets:
$q_{\mathcal{V}}(|x|)$ qubits that are private to the verifier and
$q_{\mathcal{M}}(|x|)$ qubits that represent the communication channel
between the prover and verifier.
One of the verifier's private qubits is designated as the output qubit,
which indicates whether the verifier accepts or rejects.

A quantum prover $P$ is a function mapping each input $x$ to an
$l(|x|)$-tuple $(P(x)_1,\ldots,P(x)_{l(|x|)})$ of quantum circuits.
Each of these circuits acts on $q_{\mathcal{M}}(|x|) + q_{\mathcal{P}}(|x|)$
qubits: $q_{\mathcal{P}}(|x|)$ qubits that are private to the prover and
$q_{\mathcal{M}}(|x|)$ qubits representing the communication channel.
Unlike the verifier, no restrictions are placed on the complexity of the
mapping $P$, the gates from which each $P(x)_j$ is composed, or on the size of
each $P(x)_j$, so in general we may simply view each $P(x)_j$ as an arbitrary
unitary transformation.

A verifier $V$ and a prover $P$ are compatible if for all inputs $x$ we have
(i) each $V(x)_i$ and $P(x)_j$ agree on the number $q_{\mathcal{M}}(|x|)$ of
message qubits upon which they act, and (ii)
$k(|x|)=\lfloor m(|x|)/2+1\rfloor$ and $l(|x|)=\lfloor m(|x|)/2+1/2\rfloor$
for some $m(|x|)$ (representing the number of messages exchanged).
We say that $V$ is an $m$-message verifier and $P$ is an $m$-message prover in
this case.
Whenever we discuss an interaction between a prover and verifier, we naturally
assume they are compatible.

\begin{figure}[!t]
\begin{center}
\setlength{\unitlength}{1771sp}%
\begingroup\makeatletter\ifx\SetFigFont\undefined%
\gdef\SetFigFont#1#2#3#4#5{%
  \reset@font\fontsize{#1}{#2pt}%
  \fontfamily{#3}\fontseries{#4}\fontshape{#5}%
  \selectfont}%
\fi\endgroup%
\begin{picture}(13224,5472)(589,-5623)
\thinlines
\put(1801,-3961){\framebox(1200,3000){}}
\put(6601,-3961){\framebox(1200,3000){}}
\put(11401,-3961){\framebox(1200,3000){}}
\put(601,-1261){\line( 1, 0){1200}}
\put(601,-1561){\line( 1, 0){1200}}
\put(601,-1861){\line( 1, 0){1200}}
\put(601,-1711){\line( 1, 0){1200}}
\put(601,-1411){\line( 1, 0){1200}}
\put(601,-2011){\line( 1, 0){1200}}
\put(3001,-2761){\line( 1, 0){1200}}
\put(3001,-2911){\line( 1, 0){1200}}
\put(3001,-3061){\line( 1, 0){1200}}
\put(3001,-3211){\line( 1, 0){1200}}
\put(3001,-3361){\line( 1, 0){1200}}
\put(3001,-3511){\line( 1, 0){1200}}
\put(601,-2161){\line( 1, 0){1200}}
\put(601,-1111){\line( 1, 0){1200}}
\put(3001,-3661){\line( 1, 0){1200}}
\put(3001,-3811){\line( 1, 0){1200}}
\put(4201,-5611){\framebox(1200,3000){}}
\put(9001,-5611){\framebox(1200,3000){}}
\put(3001,-1561){\line( 1, 0){3600}}
\put(3001,-1111){\line( 1, 0){3600}}
\put(3001,-1261){\line( 1, 0){3600}}
\put(3001,-1411){\line( 1, 0){3600}}
\put(3001,-1711){\line( 1, 0){3600}}
\put(3001,-1861){\line( 1, 0){3600}}
\put(3001,-2011){\line( 1, 0){3600}}
\put(3001,-2161){\line( 1, 0){3600}}
\put(7801,-1111){\line( 1, 0){3600}}
\put(7801,-1261){\line( 1, 0){3600}}
\put(7801,-1411){\line( 1, 0){3600}}
\put(7801,-1561){\line( 1, 0){3600}}
\put(7801,-1711){\line( 1, 0){3600}}
\put(7801,-1861){\line( 1, 0){3600}}
\put(7801,-2011){\line( 1, 0){3600}}
\put(7801,-2161){\line( 1, 0){3600}}
\put(601,-2761){\line( 1, 0){1200}}
\put(601,-2911){\line( 1, 0){1200}}
\put(601,-3061){\line( 1, 0){1200}}
\put(601,-3211){\line( 1, 0){1200}}
\put(601,-3361){\line( 1, 0){1200}}
\put(601,-3511){\line( 1, 0){1200}}
\put(601,-3661){\line( 1, 0){1200}}
\put(601,-3811){\line( 1, 0){1200}}
\put(5401,-2761){\line( 1, 0){1200}}
\put(5401,-2911){\line( 1, 0){1200}}
\put(5401,-3061){\line( 1, 0){1200}}
\put(5401,-3211){\line( 1, 0){1200}}
\put(5401,-3361){\line( 1, 0){1200}}
\put(5401,-3511){\line( 1, 0){1200}}
\put(5401,-3661){\line( 1, 0){1200}}
\put(5401,-3811){\line( 1, 0){1200}}
\put(7801,-2761){\line( 1, 0){1200}}
\put(7801,-2911){\line( 1, 0){1200}}
\put(7801,-3061){\line( 1, 0){1200}}
\put(7801,-3211){\line( 1, 0){1200}}
\put(7801,-3361){\line( 1, 0){1200}}
\put(7801,-3511){\line( 1, 0){1200}}
\put(7801,-3661){\line( 1, 0){1200}}
\put(7801,-3811){\line( 1, 0){1200}}
\put(10201,-2761){\line( 1, 0){1200}}
\put(10201,-2911){\line( 1, 0){1200}}
\put(10201,-3061){\line( 1, 0){1200}}
\put(10201,-3211){\line( 1, 0){1200}}
\put(10201,-3361){\line( 1, 0){1200}}
\put(10201,-3511){\line( 1, 0){1200}}
\put(10201,-3661){\line( 1, 0){1200}}
\put(10201,-3811){\line( 1, 0){1200}}
\put(12601,-1111){\line( 1, 0){1200}}
\put(12601,-1261){\line( 1, 0){1200}}
\put(12601,-1411){\line( 1, 0){1200}}
\put(12601,-1561){\line( 1, 0){1200}}
\put(12601,-1711){\line( 1, 0){1200}}
\put(12601,-1861){\line( 1, 0){1200}}
\put(12601,-2161){\line( 1, 0){1200}}
\put(12601,-2011){\line( 1, 0){1200}}
\put(12601,-2761){\line( 1, 0){1200}}
\put(12601,-2911){\line( 1, 0){1200}}
\put(12601,-3061){\line( 1, 0){1200}}
\put(12601,-3211){\line( 1, 0){1200}}
\put(12601,-3361){\line( 1, 0){1200}}
\put(12601,-3511){\line( 1, 0){1200}}
\put(12601,-3661){\line( 1, 0){1200}}
\put(12601,-3811){\line( 1, 0){1200}}
\put(601,-5461){\line( 1, 0){3600}}
\put(601,-5311){\line( 1, 0){3600}}
\put(601,-5161){\line( 1, 0){3600}}
\put(601,-5011){\line( 1, 0){3600}}
\put(601,-4861){\line( 1, 0){3600}}
\put(601,-4711){\line( 1, 0){3600}}
\put(601,-4561){\line( 1, 0){3600}}
\put(601,-4411){\line( 1, 0){3600}}
\put(5401,-4411){\line( 1, 0){3600}}
\put(5401,-4561){\line( 1, 0){3600}}
\put(5401,-4711){\line( 1, 0){3600}}
\put(5401,-4861){\line( 1, 0){3600}}
\put(5401,-5011){\line( 1, 0){3600}}
\put(5401,-5161){\line( 1, 0){3600}}
\put(5401,-5311){\line( 1, 0){3600}}
\put(5401,-5461){\line( 1, 0){3600}}
\put(10201,-4411){\line( 1, 0){3600}}
\put(10201,-4561){\line( 1, 0){3600}}
\put(10201,-4711){\line( 1, 0){3600}}
\put(10201,-4861){\line( 1, 0){3600}}
\put(10201,-5011){\line( 1, 0){3600}}
\put(10201,-5161){\line( 1, 0){3600}}
\put(10201,-5311){\line( 1, 0){3600}}
\put(10201,-5461){\line( 1, 0){3600}}
\put(1951,-2461){$V_1(x)$}
\put(6751,-2461){$V_2(x)$}
\put(11551,-2461){$V_3(x)$}
\put(4326,-4111){$P_1(x)$}
\put(9126,-4111){$P_2(x)$}

\put(-1400,-1725){\parbox{2cm}{\begin{center}\small verifier's\\private\\qubits
\end{center}}}

\put(-1400,-3425){\parbox{2cm}{\begin{center}\small message\\qubits
\end{center}}}

\put(-1400,-5025){\parbox{2cm}{\begin{center}\small prover's\\private\\qubits
\end{center}}}

\put(14000,-1200){$\leftarrow$ \hspace{-6mm}
\parbox{2cm}{\begin{center}\small output\\
qubit\end{center}}}

\end{picture}
\end{center}
\caption{Quantum circuit for a 4-message quantum interactive proof system}
\label{fig:QIP_circuit}
\end{figure}
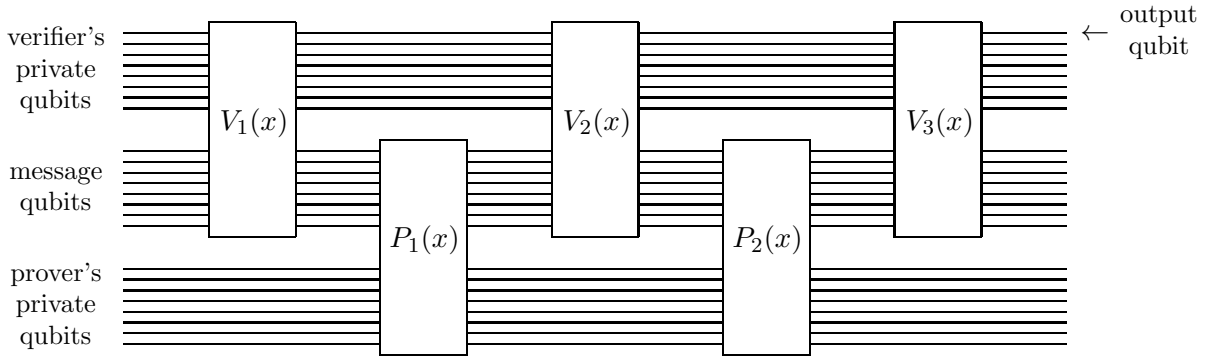

Given a verifier $V$, a prover $P$, and an input $x$, we define a quantum
circuit $(V(x),P(x))$ acting on
$q(|x|) = q_{\mathcal{V}}(|x|)+q_{\mathcal{M}}(|x|)+q_{\mathcal{P}}(|x|)$
qubits as follows.
If $m(|x|)$ is even, circuits
\[
V(x)_1,\:P(x)_1,\:\ldots,\:P(x)_{m(|x|)/2},\: V(x)_{m(|x|)/2+1}
\]
are applied in sequence, each to the
$q_{\mathcal{V}}(|x|)+q_{\mathcal{M}}(|x|)$ verifier/message qubits
or to the
$q_{\mathcal{M}}(|x|)+q_{\mathcal{P}}(|x|)$ message/prover qubits
accordingly.
This situation is illustrated in Figure~\ref{fig:QIP_circuit} for the case
$m(|x|) = 4$.
If $m(|x|)$ is odd the situation is similar, except that the prover
applies the first circuit, so circuits
\[
P(x)_1,\: V(x)_1,\: \ldots,\: P(x)_{(m(|x|)+1)/2},\: V(x)_{(m(|x|)+1)/2}
\]
are applied in sequence.
Thus, it is assumed that the prover always sends the last message (since
there would be no point for the verifier to send a message without a response).

Now, for a given input $x$, the probability that the pair $(V,P)$ accepts $x$
is defined to be the probability that an observation of the verifier's output
qubit (in the $\{\ket{0},\ket{1}\}$ basis) yields the value 1, after the
circuit $(V(x),P(x))$ is applied to a collection of $q(|x|)$ qubits each
initially in the $\ket{0}$ state.
We define a function $\mathit{max\_accept}(V(x))$ (the maximum acceptance
probability of $V(x)$) to be the probability that $(V,P)$ accepts $x$
maximized over all possible $m$-message provers $P$.

A language $A$ is said to have an $m$-message quantum interactive proof
system with completeness error $\varepsilon_c$ and soundness error
$\varepsilon_s$, where $\varepsilon_c$ and $\varepsilon_s$ may be functions
of the input length, if the exists an $m$-message verifier $V$ such that
\begin{enumerate}
\item[(i)]
if $x\in A$ then $\mathit{max\_accept}(V(x))\geq 1 - \varepsilon_c(|x|)$, and
\item[(ii)]
if $x\not\in A$ then $\mathit{max\_accept}(V(x))\leq \varepsilon_s(|x|)$.
\end{enumerate}
We also say that $(V,P)$ is a quantum interactive proof system for $A$ with
completeness error $\varepsilon_c$ and soundness error $\varepsilon_s$ if $V$
satisfies these properties and $P$ is a prover that succeeds in convincing
$V$ to accept with probability at least $1-\varepsilon_c(|x|)$ when $x\in A$.

The following conventions will be used when discussing quantum interactive
proof systems.
Assume we have a prover $P$, a verifier $V$, and an input $x$.
For readability we generally drop the arguments $x$ and $|x|$ in the various
functions above when it is understood (e.g., we write $V_j$ and $P_j$ to
denote $V(x)_j$ and $P(x)_j$ for each $j$, and we write $m$ to denote
$m(|x|)$).
We let $\mathcal{V}$, $\mathcal{M}$, and $\mathcal{P}$ denote the Hilbert
spaces corresponding to the verifier's qubits, the message qubits, and the
prover's qubits, respectively.
At a given instant, the state of the qubits in the circuit $(V,P)$ is thus a
unit vector in the space $\mathcal{V}\otimes\mathcal{M}\otimes\mathcal{P}$.
Throughout this paper, we assume that operators acting on subsystems of a
given system are extended to the entire system by tensoring with the
identity.
For instance, for a 4-message proof system as illustrated in
Figure~\ref{fig:QIP_circuit}, the state of the system after all
circuits have been applied is $V_3\,P_2\,V_2\,P_1\,V_1\,\ket{0^q}$.

\subsection{(Honest verifier) quantum statistical zero-knowledge}

Now we discuss the zero-knowledge property for quantum interactive proofs.
A short discussion of our definition follows in subsection~\ref{sec:notes}.

In the classical case, the zero-knowledge property concerns the distribution
of possible conversations between the prover and verifier from the verifier's
point of view.
In the quantum case, we cannot consider the verifier's view of the
{\em entire} interaction in terms of a single quantum state in any physically
meaningful way (this issue is discussed in subsection~\ref{sec:notes} below),
so instead we consider the mixed quantum state of the verifier's private
qubits together with the message qubits at various times during the protocol.
This gives a reasonably natural way of characterizing the verifier's view of
the interaction.

It will be sufficient to consider the verifier's view after each message
is sent (since the verifier's views at all other times are easily obtained
from the views after each message is sent by running the verifier's circuits).
The zero-knowledge property will be that the mixed states representing
the verifier's view after each message is sent should be approximable to
within negligible trace distance by a polynomial-size (uniformly generated)
quantum circuit on accepted inputs.
We formalize this notion presently.

First, given a collection $\{\rho_y\}$ of mixed states, let us say that the
collection is {\em polynomial-time preparable} if there exists a
polynomial-time uniformly generated family $\{Q_y\}$ of quantum circuits, each
having a specified collection of output qubits, such that the following holds.
For each $y$, the state $\rho_y$ is the mixed state obtained by running
$Q_y$ with all input qubits initialized to the $\ket{0}$ state and then
tracing out all non-output qubits.

Next, given a verifier $V$ and a prover $P$, we define a function
$\op{view}_{V,P}(x,j)$ to be the mixed state of the verifier and
message qubits after $j$ messages have been sent during an execution
of the proof system on input $x$.
For example, if $j$ and $m$ (the total number of messages) are both even, then
\[
\op{view}_{V,P}(x,j) = \op{tr}_\mathcal{P}
P(x)_{j/2} V(x)_{j/2}\cdots P(x)_1 V(x)_1\ket{0^q}\bra{0^q}V(x)_1^{\dagger}
P(x)_1^{\dagger}\cdots V(x)_{j/2}^{\dagger}P(x)_{j/2}^{\dagger}.
\]
The other three cases are defined similarly.

Finally, given a verifier $V$ and a prover $P$, we say that the pair $(V,P)$ is
an {\em honest verifier quantum statistical zero-knowledge proof system}
for a language $A$ if
\begin{mylist}{\parindent}
\item[1.] $(V,P)$ is an interactive proof system for $A$, and
\item[2.] there exists a polynomial-time preparable set $\{\sigma_{x,i}\}$
such that
\[
x\in A \Rightarrow
\|\sigma_{x,i} - \op{view}_{V,P}(x,i)\|_{\mathrm{tr}} \leq \delta(|x|)
\]
for some negligible function $\delta$ (i.e., $\delta(n) < 1/p(n)$  for
sufficiently large $n$ for all polynomials $p$).
\end{mylist}
The polynomial-time preparable set $\{\sigma_{x,i}\}$ corresponds to the output
of a polynomial-time simulator.
The completeness and soundness error of an honest verifier quantum statistical
zero-knowledge proof system are determined by the underlying proof system.

Finally we define {QSZK} (honest verifier quantum statistical
zero-knowledge) to be the class of languages having honest verifier quantum
statistical zero-knowledge proof systems with completeness and soundness
error at most $1/3$.
We note that sequential repetition of honest verifier quantum statistical
zero-knowledge proof systems reduces completeness and soundness error
exponentially while preserving the zero-knowledge property.
Thus, we may equivalently define {QSZK} to be the class of languages having
honest verifier quantum statistical zero-knowledge proof systems with
completeness and soundness error at most $2^{-p(n)}$ for any chosen polynomial
$p$, or with completeness and soundness error satisfying satisfying
$(1 - \varepsilon_c(n)) \geq \varepsilon_s(n) + 1/p(n)$ for some
polynomial $p$ (assuming that $\varepsilon_c(n)$ and
$\varepsilon_s(n)$ are computable in time polynomial in $n$).

%-----------------------------------------------------------------------------%

\subsection{Notes on the definition}
\label{sec:notes}

A few notes regarding our definition are in order.
First, aside from the obvious difference of quantum vs.~classical information,
our definition differs from the standard definition for classical
honest-verifier statistical zero-knowledge in the following sense.
In the classical case, the simulator randomly outputs a transcript
representing the {\em entire interaction} between the prover and verifier,
while our definition requires only that the view of the verifier
{\em at each instant} can approximated by a simulator.
The main reason for this difference is that the notion of a transcript of a
quantum interaction is counter to the nature of quantum information---in
general, there is no physically meaningful way to define a transcript of a
quantum interaction.
For instance, if a verifier were to copy down everything it sees during an
interaction in order to produce such a transcript, this would be tantamount
to the verifier measuring everything it sees, which could spoil the properties
of the protocol.

This suggests the following question about {\em classical} honest verifier
statistical zero-knowledge: is the standard definition equivalent to a
definition that is analogous to ours (i.e., requiring only that a simulator
exists that takes as input any time $t$ and outputs something that is
statistically close to the verifier's view at time $t$).
We will not attempt to answer this question in this paper.

Thus, we cannot claim that our definition is a direct quantum analogue of
the standard classical definition.
However, rather than trying to give a direct quantum analogue of the classical
definition, or aim has been to provide a definition that (i) is clearly weaker
than any reasonable definition for (not necessarily honest verifier) quantum
statistical zero-knowledge in order to prove upper bounds on the resulting
complexity class, but strong enough to allow interesting bounds to be proved,
(ii)~satisfies the intuitive notion of honest verifier statistical
zero-knowledge, and (iii) is as simple as possible.
We certainly do not suggest that our definition is the only natural
definition for honest-verifier quantum statistical zero-knowledge.
However, our results suggest that our definition yields a complexity
class that is a natural quantum variant of classical statistical
zero-knowledge, given the similarity of the complete promise problems.

%-----------------------------------------------------------------------------%

\subsection{The quantum state distinguishability problem}

A promise problem consists of two disjoint sets
$A_\mathrm{yes},\,A_\mathrm{no}\subseteq\Sigma^{\ast}$.
The computational task associated with a promise problem is as
follows: we are given some $x\in A_\mathrm{yes}\cup A_\mathrm{no}$,
and the goal is to accept if $x\in A_\mathrm{yes}$ and to reject
if $x\in A_\mathrm{no}$.
Thus, the input is {\em promised} to be an element of
$A_\mathrm{yes}\cup A_\mathrm{no}$, with no requirement made in case
the input string is not in $A_\mathrm{yes}\cup A_\mathrm{no}$.
Ordinary decision problems are a special case of promise problem where
$A_\mathrm{yes}\cup A_\mathrm{no}=\Sigma^{\ast}$.
See Even, Selman, and Yacobi \cite{EvenSY84} for further information
on promise problems.
Our above definition for QSZK is stated in terms of decision problems, but
may be extended to promise problems in the straightforward way.

In this paper we will focus on the following promise problem,
which is parameterized by constants $\alpha$ and $\beta$ satisfying
$0\leq \alpha<\beta\leq 1$.
(We will focus on a restricted version of this problem where
$\alpha < \beta^2$.)

\vspace{2mm}
\noindent
\underline{$(\alpha,\beta)$-Quantum State Distinguishability
($(\alpha,\beta)$-QSD)}
\vspace{2mm}

\noindent
\begin{tabular}{@{}lp{5.6in}}
Input: & Quantum circuits $Q_0$ and $Q_1$, each acting on $m$ qubits
and having $k$ specified output qubits.\\[1mm]
Promise: & Letting $\rho_i$ denote the mixed state obtained by running $Q_i$
on state $\ket{0^m}$ and discarding (tracing out) the non-output qubits, for
$i=0,1$, we have either
\[
\|\rho_0 - \rho_1\|_{\mathrm{tr}} \leq \alpha
\;\;\;\;\mbox{or}\;\;\;\;
\|\rho_0 - \rho_1\|_{\mathrm{tr}} \geq \beta.
\]
\\[-4mm]
Output: & {Accept} if $\|\rho_0 - \rho_1\|_{\mathrm{tr}} \geq \beta$,
{reject} if $\|\rho_0 - \rho_1\|_{\mathrm{tr}} \leq \alpha$.\\[2mm]
\end{tabular}

%=============================================================================%

\section{Quantum SZK proofs for state distinguishability}
\label{sec:trace_distance}

In this section we discuss constructions for manipulating trace distances
of outputs of quantum circuits, then present quantum zero-knowledge protocols
for the $(\alpha,\beta)$-QSD problem and its complement that are based on
these constructions.
The conclusion will be that $(\alpha,\beta)$-QSD and its complement are
in {\sc QSZK} for any constants $\alpha$ and $\beta$ satisfying
$\alpha < \beta^2$.

%-----------------------------------------------------------------------------%

\subsection{Manipulating trace distance}

Sahai and Vadhan \cite{SahaiV00} give constructions for manipulating
the statistical distance between given polynomial-time sampleable
distributions.
These constructions generalize to the trace distance between polynomial-time
preparable mixed quantum states with essentially no changes.
The following theorem describes the main consequence of the constructions.

\begin{theorem}
\label{theorem:polarize}
Fix constants $\alpha$ and $\beta$ satisfying $0\leq\alpha<\beta^2\leq 1$.
There is a (deterministic) polynomial-time procedure that, on input
$(Q_0,Q_1,1^n)$ where $Q_0$ and $Q_1$ are descriptions of quantum
circuits specifying mixed states $\rho_0$ and $\rho_1$, outputs
descriptions of quantum circuits $(R_0,R_1)$ (each having size polynomial in
$n$ and in the size of $Q_0$ and $Q_1$) specifying mixed states $\xi_0$ and
$\xi_1$ satisfying the following.
\begin{eqnarray*}
\|\rho_0 - \rho_1\|_{\mathrm{tr}} \:<\: \alpha
& \Rightarrow & 
\|\xi_0-\xi_1\|_{\mathrm{tr}} \:<\: 2^{-n},\\
\|\rho_0-\rho_1\|_{\mathrm{tr}} \:>\: \beta
& \Rightarrow &
\|\xi_0-\xi_1\|_{\mathrm{tr}} \:>\: 1-2^{-n}.
\end{eqnarray*}
\end{theorem}

\noindent
The remainder of this subsection contains a proof of this theorem.
The proof relies on the following two lemmas.

\begin{lemma}
\label{lemma:XOR2}
There is a (deterministic) polynomial-time procedure that, on input
$(Q_0,Q_1,1^r)$ where $Q_0$ and $Q_1$ are descriptions of quantum
circuits each having $k$ specified output qubits, outputs $(R_0,R_1)$,
where $R_0$ and $R_1$ are descriptions of quantum circuits each having
$rk$ specified output qubits and satisfying the following.
Letting $\rho_0,\,\rho_1,\,\xi_0$, and $\xi_1$ denote the mixed states obtained
by running $Q_0$, $Q_1$, $R_0$, and $R_1$ with all inputs in the $\ket{0}$
state and tracing out the output qubits, we have
\[
\|\xi_0 - \xi_1\|_\mathrm{tr} \: = \: \|\rho_0 - \rho_1\|_\mathrm{tr}^r.
\]
\end{lemma}
{\bf Proof.}
The circuit $R_0$ operates as follows: choose $b_1,\ldots,b_{r-1}\in\{0,1\}$
independently and uniformly, set $b_r = b_1\oplus\cdots\oplus b_{r-1}$, and
output the state $\rho_{b_1}\otimes\cdots\otimes\rho_{b_r}$ (by running
$Q_{b_1},\ldots,Q_{b_r}$ on $r$ separate collections of $k$ qubits).
The circuit $R_1$ operates similarly, except $b_r$ is flipped: randomly
choose $b_1,\ldots,b_{r-1}\in\{0,1\}$ uniformly, set
$b_r = 1\oplus b_1\oplus\cdots\oplus b_{r-1}$, and output the
state $\rho_{b_1}\otimes\cdots\otimes\rho_{b_r}$.
In both cases, the random choices are easily implemented using the
Hadamard transform, and the construction of the circuits is straightforward.
The required inequality
$\|\xi_0 - \xi_1\|_\mathrm{tr} = \|\rho_ - \rho_1\|_\mathrm{tr}^r$
follows from Proposition~\ref{prop:XOR1} (in the appendix) along with a simple
proof by induction.
\qed

\begin{lemma}
\label{lemma:amplify2}
There is a (deterministic) polynomial-time procedure that, on input
$(Q_0,Q_1,1^r)$ where $Q_0$ and $Q_1$ are descriptions of quantum
circuits each having $k$ specified output qubits, outputs $(R_0,R_1)$,
where $R_0$ and $R_1$ are descriptions of quantum circuits each having
$rk$ specified output qubits and satisfying the following.
Letting $\rho_0,\,\rho_1,\,\xi_0$, and $\xi_1$ denote the mixed states
obtained by running $Q_0$, $Q_1$, $R_0$, and $R_1$ with all inputs in the
$\ket{0}$ state and tracing out the output qubits, we have
\[
1 - \exp\left(-\frac{r}{2}\,\|\rho_0 - \rho_1\|_\mathrm{tr}^2\right)
\:\leq\:
\|\xi_0 - \xi_1\|_\mathrm{tr}
\:\leq\:
r\,\|\rho_0 - \rho_1\|_\mathrm{tr}.
\]
\end{lemma}
{\bf Proof.}
$R_0$ and $R_1$ are each simply obtained by running $r$ independent
copies of $Q_0$ and $Q_1$, respectively.
Thus $\xi_i = \rho_i^{\otimes r}$ for $i=0,1$.
The bounds on $\|\xi_0-\xi_1\|_\mathrm{tr}$ follow from
Lemma~\ref{lemma:amplify} (in the appendix).
\qed

\noindent
{\bf Proof of Theorem~\ref{theorem:polarize}.}
We assume $Q_0$ and $Q_1$ each act on $m$ qubits and have $k$ specified output
qubits for some choice of $m$ and $k$.

Apply the construction in Lemma~\ref{lemma:XOR2} to $(Q_0,Q_1,1^r)$,
where $r = \lceil\log(8n)/\log(\beta^2/\alpha)\rceil$.
The result is circuits $Q_0'$ and $Q_1'$ that produce states $\rho_0'$ and
$\rho_1'$ satisfying
\begin{eqnarray*}
\|\rho_0 - \rho_1\|_{\mathrm{tr}} \:<\: \alpha
& \Rightarrow &
\|\rho_0' - \rho_1'\|_{\mathrm{tr}} \:<\: \alpha^r\\
\|\rho_0 - \rho_1\|_{\mathrm{tr}} \:>\: \beta
& \Rightarrow &
\|\rho_0' - \rho_1'\|_{\mathrm{tr}} \:>\: \beta^r.
\end{eqnarray*}
Now apply the construction from Lemma~\ref{lemma:amplify2} to
$(Q_0',Q_1',1^s)$,
where $s=\lfloor\alpha^{-r}/2\rfloor$.
This results in circuits $Q_0''$ and $Q_1''$ that produce $\rho_0''$ and
$\rho_1''$ such that
\begin{eqnarray*}
\|\rho_0 - \rho_1\|_{\mathrm{tr}} \:<\: \alpha
& \Rightarrow &
\|\rho_0''-\rho_1''\|_{\mathrm{tr}} \,<\, \alpha^r\alpha^{-r}/2
\,=\, 1/2,\\
\|\rho_0 - \rho_1\|_{\mathrm{tr}} \,>\, \beta
& \Rightarrow &
\|\rho_0'' - \rho_1''\|_{\mathrm{tr}} >
1 - \op{exp}\left(-\frac{s}{2}\beta^{2r}\right)
%\geq
%1 - \op{exp}\left(-\frac{1}{4}\left(\frac{\beta^2}{\alpha}\right)^r
%+ 1 \right)
\geq 1 - e^{-2n+1}.
\end{eqnarray*}
Finally, again apply the construction from Lemma~\ref{lemma:XOR2}, this time to
$(Q_0'',Q_1'',1^n)$.
This results in circuits $R_0$ and $R_1$ that produce states $\xi_0$ and
$\xi_1$ satisfying
\begin{eqnarray*}
\|\rho_0 - \rho_1\|_{\mathrm{tr}} \:<\: \alpha
& \Rightarrow & 
\|\xi_0-\xi_1\|_{\mathrm{tr}} \:<\: 2^{-n},\\
\|\rho_0-\rho_1\|_{\mathrm{tr}} \:>\: \beta
& \Rightarrow &
\|\xi_0-\xi_1\|_{\mathrm{tr}} \:>\: \left(1-e^{-2n+1}\right)^n \:>\: 1-2^{-n}.
\end{eqnarray*}
The circuits $R_0$ and $R_1$ have size polynomial in $n$ and the size
of $Q_0$ and $Q_1$ as required.
\qed

%-----------------------------------------------------------------------------%

\subsection{Distance test}

Here we describe a quantum statistical zero-knowledge protocol for
Quantum State Distinguishability.
The protocol is identical in principle to several classical zero-knowledge
protocols, including the well-known Graph Non-isomorphism protocol of
Goldreich, Micali, and Wigderson \cite{GoldreichM+91} and Quadratic
Non-residuosity protocol of Goldwasser, Micali, and Rackoff
\cite{GoldwasserM+89}.

In the present case the goal of the prover is to prove that two mixed quantum
states are far apart in the trace norm metric.
A proof system for this problem is that the verifier simply prepares one of
the two states, chosen at random, and sends it to the prover, and the prover
is challenged to identify which of the two states the verifier sent.
If the states are indeed far apart, the prover can determine which state
was sent by performing an appropriate measurement, while if the states
are close together, the prover cannot reliably tell the
difference between the states because there does not exist a measurement
that distinguishes them.
By requiring that the verifier first apply the construction from the
previous section, an exponentially small error is achieved, which makes
it very easy to prove that the zero-knowledge property holds.
A more precise description of the protocol is as follows:

\vspace{3mm}

\noindent
\begin{tabular}{lp{5in}}
\underline{Verifier}: &
Apply the construction of Theorem~\ref{theorem:polarize} to $(Q_0,Q_1,1^n)$
for $n$ exceeding the length of the input $(Q_0,Q_1)$.
Let $R_0$ and $R_1$ denote the constructed circuits, and $\xi_0$ and $\xi_1$
the associated mixed states.
Choose $b\in\{0,1\}$ uniformly and send $\xi_b$ to the prover. \\[2mm]
\underline{Honest prover}: & Perform the optimal measurement for
distinguishing $\xi_0$ and $\xi_1$.
Let $\tilde{b}$ be 0 if the measurement indicates the state is $\xi_0$,
and 1 if the measurement indicates the state is $\xi_1$.
Send $\tilde{b}$ to the verifier.\\[2mm]
\underline{Verifier}: & {\em Accept} if $b = \tilde{b}$ and {\em reject}
otherwise.
\end{tabular}
\vspace{4mm}

\noindent
Based on this protocol, we have the following theorem.

\begin{theorem}
\label{theorem:distance}
Let $\alpha$ and $\beta$ be constants satisfying $0\leq \alpha <\beta^2\leq 1$.
Then $(\alpha,\beta)$-QSD $\in$ {QSZK}.
\end{theorem}

\noindent
{\bf Proof.}
First we discuss the completeness and soundness of the proof system, then
prove that the zero-knowledge property holds.

For the completeness property of the protocol, we assume that the prover
receives one of $\xi_0$ and $\xi_1$ such that
$\|\xi_0 - \xi_1\|_\mathrm{tr}> 1 - 2^{-n}$, and thus can distinguish the two
cases with probability of error bounded by $2^{-n}$ by performing an
appropriate measurement.
Specifically, the prover can apply the measurement described by orthogonal
projections $\{\Pi_0,\Pi_1\}$ where $\Pi_0$ maximizes
$\op{tr}\Pi_0(\xi_0-\xi_1)$ and $\Pi_1 = I - \Pi_0$.
This gives an outcome of 0 with probability at least $1 - 2^{-n}$ in
case the verifier sent $\xi_0$ and gives an outcome of 1 with probability at
least $1 - 2^{-n}$ in case the verifier sent $\xi_1$.
This will cause the verifier to accept with probability at least $1-2^{-n}$.

For the soundness condition, we assume the prover receives either $\xi_0$ or
$\xi_1$ where $\|\xi_0-\xi_1\|_\mathrm{tr}<2^{-n}$, and then the prover
returns a single bit to the verifier.
There is no loss of generality in assuming that the bit sent by the prover
is measured immediately upon being received by the verifier, since this
would not change the verifier's decision to accept or reject.
Thus, we may treat this bit as being the outcome of a measurement of
whichever state $\xi_0$ or $\xi_1$ was initially sent by the verifier.
Since the trace distance between these two states is at most $2^{-n}$,
no measurement can distinguish the states with bias exceeding $2^{-n}$.
Consequently the prover has probability at most $1/2 + 2^{-n}$ of
correctly answering $\tilde{b} = b$.

Finally, the zero-knowledge property is straightforward---the state
of the verifier and message qubits after the first message is obtained
by applying $V_1$ (the verifier's first transformation), and the
state of the verifier and message qubits after the prover's response
is approximated by applying $V_1$, tracing out the message qubits,
then setting $\tilde{b}$ to $b$.
Since the completeness error is exponentially small, this gives a
negligible error for the simulator.
\qed

%-----------------------------------------------------------------------------%

\subsection{Closeness test}

Now we consider a protocol for the complement of $(\alpha,\beta)$-QSD.
Unlike the previous protocol this protocol seems to have no classical
analogue, relying heavily on non-classical properties of quantum states.

We begin with a description of the protocol, which is as follows:

\vspace{3mm}

\noindent
\begin{tabular}{lp{5in}}
\underline{Verifier}: &
Apply the construction of Theorem~\ref{theorem:polarize} to $(Q_0,Q_1,1^{n+1})$
for $n$ exceeding the length of the input $(Q_0,Q_1)$.
Let $R_0$ and $R_1$ denote the constructed circuits, and $\xi_0$ and $\xi_1$
the associated mixed states.
Let $t$ be the number of qubits on which $R_0$ and $R_1$ act.
Apply $R_0$ to $\ket{0^t}$ and send the prover {\bf only} the non-output
qubits (that is, the qubits that would be traced-out to yield $\xi_0$).\\[2mm]

\underline{Honest prover}: & Apply unitary transformation $U$ (described
below) to the qubits sent by the verifier, then send these qubits back to
the verifier.\\[2mm]

\underline{Verifier}: & Apply $R_1^{\dagger}$ to the output qubits of
$R_0$ (which were not send to the prover in the first message) together with
the qubits received from the prover.
Measure the resulting qubits: {\em accept} if the result is $0^t$, and
{\em reject} otherwise.
\end{tabular}

\vspace{3mm}

\noindent
The correctness of the protocol is closely related to the Schmidt
decomposition of bipartite quantum states, which states the following.
If $\ket{\phi}\in\mathcal{H}\otimes\mathcal{K}$ is a pure, bipartite quantum
state, then it is possible to write
\[
\ket{\phi} = \sum_{i=1}^n\sqrt{p_i}\,\ket{\psi_i}\ket{\nu_i}
\]
for positive real numbers $p_1,\ldots,p_n$ and orthonormal sets
$\{\ket{\psi_1},\ldots,\ket{\psi_n}\}$ and
$\{\ket{\nu_1},\ldots,\ket{\nu_n}\}$.
Such sets may be obtained by letting $\{\ket{\psi_1},\ldots,\ket{\psi_n}\}$
be an orthonormal collection of eigenvectors of
$\rho = \op{tr}_\mathcal{K}\ket{\phi}\bra{\phi}$ having nonzero eigenvalues
and taking $p_1,\ldots,p_n$ to be the corresponding nonzero eigenvalues,
which are therefore positive since $\rho$ is positive semidefinite.
At this point $\ket{\nu_1},\ldots,\ket{\nu_n}$ are determined, and
can be shown to be orthonormal.
Consequently, if we have two bipartite states
$\ket{\phi},\ket{\phi'}\in\mathcal{H}\otimes\mathcal{K}$ that give the same
mixed state when the second system is traced-out, i.e.,
$\op{tr}_\mathcal{K}\ket{\phi}\bra{\phi} =
\op{tr}_\mathcal{K}\ket{\phi'}\bra{\phi'} = \rho$, then there must exist a
unitary operator $U$ acting on $\mathcal{K}$ such that
$(I\otimes U)\ket{\phi} = \ket{\phi'}$.
The operator $U$ is simply a change of basis taking
$\ket{\nu_i}$ to $\ket{\nu'_i}$ for each $i$, where the vectors
$\ket{\nu_1'},\ldots,\ket{\nu_n'}$ are given by
\[
\ket{\phi'} = \sum_{i=1}^n\sqrt{p_i}\,\ket{\psi_i}\ket{\nu'_i}.
\]
In case $\rho = \op{tr}_\mathcal{K}\ket{\phi}\bra{\phi}$ and
$\rho' = \op{tr}_\mathcal{K}\ket{\phi'}\bra{\phi'}$ are not identical, but
are close together in the trace norm metric, an approximate version of
this fact holds: there exists a unitary operator $U$ acting on $\mathcal{K}$
such that $(I\otimes U)\ket{\phi}$ and $\ket{\phi'}$ are close in Euclidean
norm.
For the above protocol, the states $\ket{\phi}$ and $\ket{\phi'}$ are the
states produced by $R_0$ and $R_1$, $\mathcal{K}$ is the space corresponding
to the qubits sent to the prover, and $U$ corresponds to the action of the
prover.

We formalize this argument in the proof of the following theorem.

\begin{theorem}
\label{theorem:complement_in_QSZK}
Let $\alpha$ and $\beta$ satisfy $0\leq \alpha <\beta^2\leq 1$.
Then $(\alpha,\beta)$-QSD $\in$ {co-QSZK}.
\end{theorem}

\noindent
{\bf Proof.}
First let us consider the completeness condition.
If $(Q_0,Q_1)\not\in (\alpha,\beta)$-QSD then we have
$\|\xi_0 - \xi_1\|_\mathrm{tr}<2^{-(n+1)}$ and thus
$F(\xi_0,\xi_1)> 1 - 2^{-(n+1)}$ (where $F(\xi_0,\xi_1)$ denotes the
fidelity of $\xi_0$ and $\xi_1$).
The states $R_0\ket{0^t}$ and $R_1\ket{0^t}$ are purifications of $\xi_0$
and $\xi_1$, respectively, so by Lemma~\ref{lemma:closeness1} (in the
appendix) there exists a unitary transformation $U$ acting only on the
non-output qubits of $R_0\ket{0^t}$ (i.e., the qubits sent to the prover)
such that $\|(I\otimes U)R_0\ket{0^t} - R_1\ket{0^t}\|\leq 2^{-n/2}$.
This is the transformation $U$ performed by the honest prover.
The verifier accepts with probability
\[
|\bra{0^t}R_1^{\dagger}(I\otimes U)R_0\ket{0^t}|^2 \:\geq\:
\left(1-\frac{1}{2}\|R_1\ket{0^t}-(I\otimes U)R_0\ket{0^t}\|^2\right)^2
\:>\: 1 - 2^{-n}.
\]

The soundness of the proof system may be proved as follows.
Assume $(Q_0,Q_1)\in (\alpha,\beta)$-QSD, so that
$\|\xi_0 - \xi_1\|_\mathrm{tr}> 1 - 2^{-(n+1)}$, and thus
$F(\xi_0,\xi_1)< 2^{-n/2}$.
The verifier prepares $R_0\ket{0^t}$ and sends the non-output qubits
to the prover.
The most general action of the prover is to apply some arbitrary unitary
transformation to the qubits sent by the verifier along with any number
of its own private qubits, and then return some number of these qubits to the
verifier.
Let $\sigma$ denote the mixed state of the verifier's private qubits
and the message qubits immediately after the prover has sent its message.
As usual we let $\mathcal{V}$ denote the space corresponding to the verifier's
private qubits and $\mathcal{M}$ the space corresponding to the message
qubits, so that $\sigma\in\mathbf{D}(\mathcal{V}\otimes\mathcal{M})$ and
$\op{tr}_\mathcal{M}\sigma = \xi_0$.
(The fact that $\op{tr}_\mathcal{M}\sigma = \xi_0$ follows from the
fact that the prover has not touched the verifier's private qubits, so
that they must still be in state $\xi_0$.)
The verifier applies $R_1^{\dagger}$ and measures, which results in
{\em accept} with probability $\bra{0^t}R_1^{\dagger}\sigma R_1\ket{0^t}$.
Since $R_1\ket{0^t}$ is a purification of $\xi_1$, we have that
$\bra{0^t}R_1^{\dagger}\sigma R_1\ket{0^t} \leq F(\xi_0,\xi_1)^2 < 2^{-n}$
by Lemma~\ref{cor:of_Uhlmann} (in the appendix).
Thus the verifier accepts with exponentially small probability.

Finally, the zero-knowledge property is again straightforward.
We define a simulator that outputs $R_0\ket{0^t}$ for the verifier's view
as the first message is being sent and $R_1\ket{0^t}$ for the verifier's view
after the second message.
The simulator is perfect for the first message, and has trace
distance at most $2^{-n}$ from the actual view of the verifier interacting
with the prover defined above for the second message.
\qed

%=============================================================================%

\section{Completeness of quantum state distinguishability for QSZK}
\label{sec:completeness}

The notion of a promise problem being complete for a given class is
defined in the most straightforward way; in the case of QSZK we say
that a promise problem $B = (B_\mathrm{yes},B_\mathrm{no})$ is complete for
QSZK if (i) $B \in \mbox{QSZK}$, and (ii) for every promise problem
$A = (A_\mathrm{yes},A_\mathrm{no})\in\mbox{QSZK}$ there is a
deterministic polynomial-time computable function $f$ such that
for all $x$ we have
$x\in A_\mathrm{yes}\;\Rightarrow\;f(x)\in B_\mathrm{yes}$ and
$x\in A_\mathrm{no}\;\Rightarrow\;f(x)\in B_\mathrm{no}$.
In this section we prove that $(\alpha,\beta)$-QSD is complete for QSZK
whenever $\alpha$ and $\beta$ are constants satisfying
$0 < \alpha < \beta^2 < 1$.

\begin{theorem}
\label{theorem:complete}
Let $\alpha$ and $\beta$ satisfy $0<\alpha<\beta^2<1$.
Then $(\alpha,\beta)$-QSD is complete for {QSZK}.
\end{theorem}

\noindent
By Theorems~\ref{theorem:distance} and \ref{theorem:complement_in_QSZK} we
have that $(\alpha,\beta)$-QSD is in {QSZK} $\cap$ co-QSZK provided
$\alpha < \beta^2$.
In order to prove Theorem~\ref{theorem:complete} it will therefore suffice to
show that, for any promise problem $A\in\mbox{QSZK}$, $A$ reduces to the
complement of \mbox{$(\alpha,\beta)$-QSD}.
After describing the reduction $f$, the main facts to be proved will therefore
be
\begin{enumerate}
\item[(i)] $x\in A_\mathrm{yes}\;\Rightarrow\;f(x)\in
(\alpha,\beta)\mbox{-QSD}_\mathrm{no}$, and
\item[(ii)] $x\in A_\mathrm{no}\;\Rightarrow\;f(x)\in
(\alpha,\beta)\mbox{-QSD}_\mathrm{yes}$.
\end{enumerate}

\noindent
The following technical lemma will be useful in the proof.

\begin{lemma}
\label{lemma:complete1}
Let $V$ be an $m$-message verifier and $x$ an input such that $m = m(|x|)$ is
even and\linebreak
$\mathit{max\_accept}(V(x))\leq\varepsilon$.
Let $k = m/2 +1$, so that $V(x) = (V_1,\ldots,V_k)$.
Let $\rho_0,\ldots,\rho_{k-1}\in\mathbf{D}(\mathcal{V}\otimes\mathcal{M})$,
let $\xi_i = V_i\rho_{i-1}V_i^{\dagger}$ for $i=1,\ldots,k$, and assume
that $\rho_0 = \ket{0^{q_{\mathcal{V}} + q_{\mathcal{M}}}}
\bra{0^{q_{\mathcal{V}} + q_{\mathcal{M}}}}$
(i.e., $\rho_0$ denotes the initial state of the qubits)
and $\op{tr}(\Pi_\mathrm{acc}\,\xi_k) = 1$
for $\Pi_\mathrm{acc}$ denoting the projection onto states for which
the output qubit has value 1 (i.e., $\xi_k$ is a state where the verifier
accepts with certainty).
Then
\[
\|\op{tr}_{\mathcal{M}}\xi_1\otimes\cdots\otimes\op{tr}_{\mathcal{M}}\xi_{k-1}
-\op{tr}_{\mathcal{M}}\rho_1\otimes\cdots\otimes\op{tr}_{\mathcal{M}}\rho_{k-1}
\|_{\mathrm{tr}} \:\geq\:\frac{(1 - \sqrt{\varepsilon})^2}{3(k-1)}.
\]
\end{lemma}

\noindent
{\bf Proof.}
Let $\ket{\phi_0} = \ket{0^q}$, which is a purification of $\rho_0$,
let $\ket{\phi_1},\ldots,\ket{\phi_{k-1}}\in\mathcal{V}\otimes\mathcal{M}
\otimes\mathcal{P}$ be purifications of $\rho_1,\ldots,\rho_{k-1}$, and
set $\ket{\psi_i} = V_i\ket{\phi_{i-1}}$ for $i = 1,\ldots,k$.
(As usual, we extend each $V_i$ to a unitary operator on
$\mathcal{V}\otimes\mathcal{M}\otimes\mathcal{P}$ by tensoring with the
identity on $\mathcal{P}$).
Note that $\ket{\psi_1},\ldots,\ket{\psi_k}$ are necessarily purifications
of $\xi_1,\ldots,\xi_k$.

Define
\[
\delta_i\:=\: 1 - F(\op{tr}_{\mathcal{M}}\xi_i, \op{tr}_{\mathcal{M}}\rho_i)
\]
for $i=1,\ldots,k-1$.
By Lemma~\ref{lemma:closeness1} (in the appendix) there exists a unitary
operator $P_i\in\mathbf{U}(\mathcal{M}\otimes\mathcal{P})$ such that
\[
\|P_i\ket{\psi_i} - \ket{\phi_i}\|\:\leq\:\sqrt{2\delta_i}.
\]
Now, for each $i=2,\ldots,k$, we have
\begin{eqnarray*}
\lefteqn{\|V_i P_{i-1} V_{i-1}\cdots P_1 V_1\ket{\phi_0} - \ket{\psi_i}\|}
\hspace{10mm}\\
& = & 
\|P_{i-1} V_{i-1}\cdots P_1 V_1\ket{\phi_0} - \ket{\phi_{i-1}}\|\\
& \leq &
\|P_{i-1} V_{i-1}\cdots P_1 V_1\ket{\phi_0} - P_{i-1}\ket{\psi_{i-1}}\|
+ \|P_{i-1}\ket{\psi_{i-1}} - \ket{\phi_{i-1}} \|\\
& \leq & 
\|V_{i-1}\cdots P_1 V_1\ket{\phi_0} - \ket{\psi_{i-1}}\|
+ \sqrt{2\delta_{i-1}},
\end{eqnarray*}
so that
\[
\|V_k P_{k-1} V_{k-1}\cdots P_1 V_1\ket{\phi_0} - \ket{\psi_k}\|
\:\leq\: \sum_{i=1}^{k-1}\sqrt{2\delta_i}.
\]
Consequently, since $\|\Pi_\mathrm{acc}\,\ket{\psi_k}\| = 1$, we must
have
\[
\|\Pi_\mathrm{acc}\,V_k P_{k-1}V_{k-1}\cdots P_1 V_1\ket{\phi_0}\|
\geq 1 - \sum_{i=1}^{k-1}\sqrt{2 \delta_i}.
\]
Since $\mathit{max\_accept}(V(x))\leq \varepsilon$ and $\ket{\phi_0}$ is the
initial state of $(V(x),P(x))$, this implies
\begin{equation}
\label{eq:MAP_distance1}
\sum_{i=1}^{k-1}\sqrt{2\delta_i}\:\geq\: 1 - \sqrt{\varepsilon}.
\end{equation}

Now, by Proposition~\ref{prop:fidelity_mult}, we have
\[
F(\op{tr}_{\mathcal{M}}\xi_1\otimes\cdots\otimes\op{tr}_{\mathcal{M}}\xi_{k-1},
\op{tr}_{\mathcal{M}}\rho_1\otimes\cdots\otimes\op{tr}_{\mathcal{M}}\rho_{k-1})
\: = \:
\prod_{i=1}^{k-1}F(\op{tr}_{\mathcal{M}}\xi_i,\op{tr}_{\mathcal{M}}\rho_i)
\: \leq \:
\prod_{i=1}^{k-1}(1 - \delta_i).
\]
Subject to the constraint in Eq.~\ref{eq:MAP_distance1}, we have
\[
\prod_{i=1}^{k-1}(1 - \delta_i) \: \leq \:
\left(1 - \frac{(1 - \sqrt{\varepsilon})^2}{2(k-1)^2}\right)^{k-1}\:\leq\:
\op{exp}\left(-\frac{(1 - \sqrt{\varepsilon})^2}{2(k-1)}\right)\:\leq\:
1 - \frac{(1 - \sqrt{\varepsilon})^2}{3(k-1)}.
\]
Thus,
\[
\|\op{tr}_{\mathcal{M}}\xi_1\otimes\cdots\otimes\op{tr}_{\mathcal{M}}\xi_{k-1}
-\op{tr}_{\mathcal{M}}\rho_1\otimes\cdots\otimes\op{tr}_{\mathcal{M}}\rho_{k-1}
\|_{\mathrm{tr}} \:\geq\: \frac{(1 - \sqrt{\varepsilon})^2}{3(k-1)}
\]
as required.
\qed

\noindent{\bf Proof of Theorem~\ref{theorem:complete}.}
Let $A\in$ {QSZK}, and let $(V,P)$ be an honest verifier quantum statistical
zero-knowledge proof system for $A$ with completeness and soundness error
smaller than $2^{-n}$ for inputs of length $n$.
Such a proof system exists, since sequential repetition reduces completeness
and soundness errors exponentially while preserving the zero-knowledge
property of honest verifier quantum statistical zero-knowledge proof systems.
Let $m = m(|x|)$ be the number of messages exchanged by $P$ and $V$.
For simplicity we assume that the number of messages $m$ is even for all $x$
(adding an initial move where the verifier sends some arbitrary state if
necessary).
Thus, the verifier will apply transformations $V_1,\ldots,V_k$ for 
$k = m/2 + 1$, and will send the first message in the protocol.
We let $\{\sigma_{x,j}\}$ correspond to the mixed states output by the
simulator for $(V,P)$ as discussed in Section~\ref{sec:preliminaries}.
The quantum circuits that produce the states $\{\sigma_{x,j}\}$ are used
implicitly in the reduction.

First, we describe, for any fixed input $x$, the following quantum states:
\vspace{2mm}

\begin{tabular}{@{}lp{5.7in}}
1. & Let $\rho_0$ be the state in which all verifier and message qubits are
in state $\ket{0}$.\\[2mm]
2. &  Let $\xi_k$ denote the state obtained by applying $V_k$ to
$\sigma_{x,m}$, discarding the output qubit, and replacing it with a qubit in
state $\ket{1}$.\\[2mm]
3. & Let $\rho_i = \sigma_{x,2i}$ for $i = 1,\ldots, k-2$ and
let $\rho_{k-1} = V_k^{\dagger}\xi_k V_k$.\\[2mm]
4. & Let $\xi_i = V_i\rho_{i-1} V_i^{\dagger}$ for $i = 1,\ldots, k-1$.\\[2mm]
\end{tabular}
\vspace{2mm}

\begin{figure}[t]
\begin{center}
\setlength{\unitlength}{1871sp}%
\begingroup\makeatletter\ifx\SetFigFont\undefined%
\gdef\SetFigFont#1#2#3#4#5{%
  \reset@font\fontsize{#1}{#2pt}%
  \fontfamily{#3}\fontseries{#4}\fontshape{#5}%
  \selectfont}%
\fi\endgroup%
\begin{picture}(13224,5472)(589,-5623)
\thinlines
\put(1801,-3961){\framebox(1200,3000){}}
\put(6601,-3961){\framebox(1200,3000){}}
\put(11401,-3961){\framebox(1200,3000){}}
\put(601,-1261){\line( 1, 0){1200}}
\put(601,-1561){\line( 1, 0){1200}}
\put(601,-1861){\line( 1, 0){1200}}
\put(601,-1711){\line( 1, 0){1200}}
\put(601,-1411){\line( 1, 0){1200}}
\put(601,-2011){\line( 1, 0){1200}}
\put(3001,-2761){\line( 1, 0){1200}}
\put(3001,-2911){\line( 1, 0){1200}}
\put(3001,-3061){\line( 1, 0){1200}}
\put(3001,-3211){\line( 1, 0){1200}}
\put(3001,-3361){\line( 1, 0){1200}}
\put(3001,-3511){\line( 1, 0){1200}}
\put(601,-2161){\line( 1, 0){1200}}
\put(601,-1111){\line( 1, 0){1200}}
\put(3001,-3661){\line( 1, 0){1200}}
\put(3001,-3811){\line( 1, 0){1200}}
\put(4201,-5611){\framebox(1200,3000){}}
\put(9001,-5611){\framebox(1200,3000){}}
\put(3001,-1561){\line( 1, 0){3600}}
\put(3001,-1111){\line( 1, 0){3600}}
\put(3001,-1261){\line( 1, 0){3600}}
\put(3001,-1411){\line( 1, 0){3600}}
\put(3001,-1711){\line( 1, 0){3600}}
\put(3001,-1861){\line( 1, 0){3600}}
\put(3001,-2011){\line( 1, 0){3600}}
\put(3001,-2161){\line( 1, 0){3600}}
\put(7801,-1111){\line( 1, 0){3600}}
\put(7801,-1261){\line( 1, 0){3600}}
\put(7801,-1411){\line( 1, 0){3600}}
\put(7801,-1561){\line( 1, 0){3600}}
\put(7801,-1711){\line( 1, 0){3600}}
\put(7801,-1861){\line( 1, 0){3600}}
\put(7801,-2011){\line( 1, 0){3600}}
\put(7801,-2161){\line( 1, 0){3600}}
\put(601,-2761){\line( 1, 0){1200}}
\put(601,-2911){\line( 1, 0){1200}}
\put(601,-3061){\line( 1, 0){1200}}
\put(601,-3211){\line( 1, 0){1200}}
\put(601,-3361){\line( 1, 0){1200}}
\put(601,-3511){\line( 1, 0){1200}}
\put(601,-3661){\line( 1, 0){1200}}
\put(601,-3811){\line( 1, 0){1200}}
\put(5401,-2761){\line( 1, 0){1200}}
\put(5401,-2911){\line( 1, 0){1200}}
\put(5401,-3061){\line( 1, 0){1200}}
\put(5401,-3211){\line( 1, 0){1200}}
\put(5401,-3361){\line( 1, 0){1200}}
\put(5401,-3511){\line( 1, 0){1200}}
\put(5401,-3661){\line( 1, 0){1200}}
\put(5401,-3811){\line( 1, 0){1200}}
\put(7801,-2761){\line( 1, 0){1200}}
\put(7801,-2911){\line( 1, 0){1200}}
\put(7801,-3061){\line( 1, 0){1200}}
\put(7801,-3211){\line( 1, 0){1200}}
\put(7801,-3361){\line( 1, 0){1200}}
\put(7801,-3511){\line( 1, 0){1200}}
\put(7801,-3661){\line( 1, 0){1200}}
\put(7801,-3811){\line( 1, 0){1200}}
\put(10201,-2761){\line( 1, 0){1200}}
\put(10201,-2911){\line( 1, 0){1200}}
\put(10201,-3061){\line( 1, 0){1200}}
\put(10201,-3211){\line( 1, 0){1200}}
\put(10201,-3361){\line( 1, 0){1200}}
\put(10201,-3511){\line( 1, 0){1200}}
\put(10201,-3661){\line( 1, 0){1200}}
\put(10201,-3811){\line( 1, 0){1200}}
\put(12601,-1111){\line( 1, 0){1200}}
\put(12601,-1261){\line( 1, 0){1200}}
\put(12601,-1411){\line( 1, 0){1200}}
\put(12601,-1561){\line( 1, 0){1200}}
\put(12601,-1711){\line( 1, 0){1200}}
\put(12601,-1861){\line( 1, 0){1200}}
\put(12601,-2161){\line( 1, 0){1200}}
\put(12601,-2011){\line( 1, 0){1200}}
\put(12601,-2761){\line( 1, 0){1200}}
\put(12601,-2911){\line( 1, 0){1200}}
\put(12601,-3061){\line( 1, 0){1200}}
\put(12601,-3211){\line( 1, 0){1200}}
\put(12601,-3361){\line( 1, 0){1200}}
\put(12601,-3511){\line( 1, 0){1200}}
\put(12601,-3661){\line( 1, 0){1200}}
\put(12601,-3811){\line( 1, 0){1200}}
\put(601,-5461){\line( 1, 0){3600}}
\put(601,-5311){\line( 1, 0){3600}}
\put(601,-5161){\line( 1, 0){3600}}
\put(601,-5011){\line( 1, 0){3600}}
\put(601,-4861){\line( 1, 0){3600}}
\put(601,-4711){\line( 1, 0){3600}}
\put(601,-4561){\line( 1, 0){3600}}
\put(601,-4411){\line( 1, 0){3600}}
\put(5401,-4411){\line( 1, 0){3600}}
\put(5401,-4561){\line( 1, 0){3600}}
\put(5401,-4711){\line( 1, 0){3600}}
\put(5401,-4861){\line( 1, 0){3600}}
\put(5401,-5011){\line( 1, 0){3600}}
\put(5401,-5161){\line( 1, 0){3600}}
\put(5401,-5311){\line( 1, 0){3600}}
\put(5401,-5461){\line( 1, 0){3600}}
\put(10201,-4411){\line( 1, 0){3600}}
\put(10201,-4561){\line( 1, 0){3600}}
\put(10201,-4711){\line( 1, 0){3600}}
\put(10201,-4861){\line( 1, 0){3600}}
\put(10201,-5011){\line( 1, 0){3600}}
\put(10201,-5161){\line( 1, 0){3600}}
\put(10201,-5311){\line( 1, 0){3600}}
\put(10201,-5461){\line( 1, 0){3600}}
\put(1156,-4006){\oval(210,210)[bl]}
\put(1156,-916){\oval(210,210)[tl]}
\put(1246,-4006){\oval(210,210)[br]}
\put(1246,-916){\oval(210,210)[tr]}
\put(1156,-4111){\line( 1, 0){ 90}}
\put(1156,-811){\line( 1, 0){ 90}}
\put(1051,-4006){\line( 0, 1){3090}}
\put(1351,-4006){\line( 0, 1){3090}}
\put(3556,-4006){\oval(210,210)[bl]}
\put(3556,-916){\oval(210,210)[tl]}
\put(3646,-4006){\oval(210,210)[br]}
\put(3646,-916){\oval(210,210)[tr]}
\put(3556,-4111){\line( 1, 0){ 90}}
\put(3556,-811){\line( 1, 0){ 90}}
\put(3451,-4006){\line( 0, 1){3090}}
\put(3751,-4006){\line( 0, 1){3090}}
\put(5956,-4006){\oval(210,210)[bl]}
\put(5956,-916){\oval(210,210)[tl]}
\put(6046,-4006){\oval(210,210)[br]}
\put(6046,-916){\oval(210,210)[tr]}
\put(5956,-4111){\line( 1, 0){ 90}}
\put(5956,-811){\line( 1, 0){ 90}}
\put(5851,-4006){\line( 0, 1){3090}}
\put(6151,-4006){\line( 0, 1){3090}}
\put(8356,-4006){\oval(210,210)[bl]}
\put(8356,-916){\oval(210,210)[tl]}
\put(8446,-4006){\oval(210,210)[br]}
\put(8446,-916){\oval(210,210)[tr]}
\put(8356,-4111){\line( 1, 0){ 90}}
\put(8356,-811){\line( 1, 0){ 90}}
\put(8251,-4006){\line( 0, 1){3090}}
\put(8551,-4006){\line( 0, 1){3090}}
\put(10756,-4006){\oval(210,210)[bl]}
\put(10756,-916){\oval(210,210)[tl]}
\put(10846,-4006){\oval(210,210)[br]}
\put(10846,-916){\oval(210,210)[tr]}
\put(10756,-4111){\line( 1, 0){ 90}}
\put(10756,-811){\line( 1, 0){ 90}}
\put(10651,-4006){\line( 0, 1){3090}}
\put(10951,-4006){\line( 0, 1){3090}}
\put(13156,-4006){\oval(210,210)[bl]}
\put(13156,-916){\oval(210,210)[tl]}
\put(13246,-4006){\oval(210,210)[br]}
\put(13246,-916){\oval(210,210)[tr]}
\put(13156,-4111){\line( 1, 0){ 90}}
\put(13156,-811){\line( 1, 0){ 90}}
\put(13051,-4006){\line( 0, 1){3090}}
\put(13351,-4006){\line( 0, 1){3090}}
\put(1076,-511){$\rho_0$}
\put(3501,-511){$\xi_1$}
\put(5876,-511){$\rho_1$}
\put(8276,-511){$\xi_2$}
\put(10651,-511){$\rho_2$}
\put(13076,-511){$\xi_3$}
\put(2251,-2461){$V_1$}
\put(7051,-2461){$V_2$}
\put(11851,-2461){$V_3$}
\put(4626,-4111){$P_1$}
\put(9426,-4111){$P_2$}
\end{picture}
\end{center}
\caption{States $\rho_0,\ldots,\rho_{k-1}$ and $\xi_1,\ldots,\xi_k$
for $m = 4$, $k = 3$.}
\label{fig:states}
\end{figure}
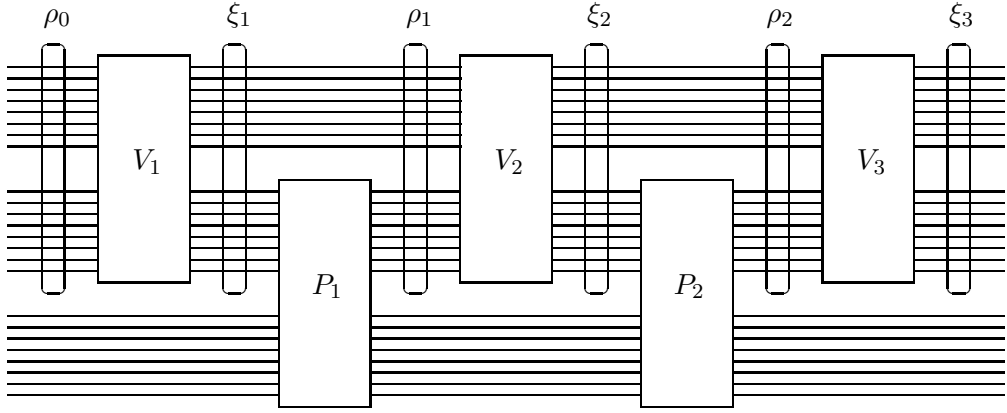

\noindent
These states are illustrated in Figure~\ref{fig:states} for the case $m = 4$
(meaning that these states will be close approximations to the
illustrated states given an input $x\in A_\mathrm{yes}$).
Let $Q_0 = Q_0(x)$ and $Q_1 = Q_1(x)$ be quantum circuits that output
$\gamma_0 = \op{tr}_\mathcal{M}(\rho_1)\otimes\cdots\otimes
\op{tr}_\mathcal{M}(\rho_{k-1})$ and
$\gamma_1 = \op{tr}_\mathcal{M}(\xi_1)\otimes\cdots\otimes
\op{tr}_\mathcal{M}(\xi_{k-1})$,
respectively (assuming the circuits are applied to the state $\ket{0^m}$
for appropriate $m$ and non-output qubits are traced out in the usual way).
These circuits are easily constructed based on $V$ and on the simulator
for $(V,P)$.

We claim that the following implications hold:
\[
x\in A_\mathrm{yes} \Rightarrow \|\gamma_0 - \gamma_1\|_\mathrm{tr}
< \delta(|x|)\;\;\;\;\;\mbox{and}\;\;\;\;\;
x\in A_\mathrm{no} \Rightarrow \|\gamma_0 - \gamma_1\|_\mathrm{tr}
> c/k
\]
where $\delta(|x|)$ is a negligible function (determined by the accuracy
of the simulator for $(V,P)$) and $c>0$ is constant.
The second implication follows immediately from Lemma~\ref{lemma:complete1}.
To prove the first implication, consider states
$\rho_0',\ldots,\rho_{k-1}'$ and $\xi_1',\ldots,\xi_k'$ obtained
precisely as in the description of $Q_0$ and $Q_1$, except replacing
$\sigma_{x,j}$ with $\op{view}_{V,P}(x,j)$, the actual view of the verifier $V$
while interacting with $P$, for each $x$ and $j$.
We necessarily have $\op{tr}_{\mathcal{M}}\xi_i'=\op{tr}_{\mathcal{M}}\rho_i'$
for $i = 1,\ldots, k-2$.
Since measuring the output qubit of $V_k\op{view}_{V,P}(x,m)V_k^{\dagger}$
gives 1 with probability at least $1 - 2^{-|x|}$, replacing the output qubit
with a qubit in state $\ket{1}$ has little effect on this state.
Specifically, we deduce that
$\|\op{tr}_\mathcal{M}\xi_{k-1}'-\op{tr}_\mathcal{M}\rho_{k-1}'\|_\mathrm{tr}
< 2^{-|x|/2}$.
Thus, the quantity
\[
\|\op{tr}_\mathcal{M}(\rho_1')\otimes\cdots\otimes
\op{tr}_\mathcal{M}(\rho_{k-1}')\: - \:\op{tr}_\mathcal{M}(\xi_1')\otimes
\cdots\otimes\op{tr}_\mathcal{M}(\xi_{k-1}')\|_\mathrm{tr}
\]
is negligible.
Now, since the simulator deviates from $\op{view}_{V,P}$ by a
negligible quantity on each input, the inequality
\[
\|\op{tr}_\mathcal{M}(\rho_1)\otimes\cdots\otimes
\op{tr}_\mathcal{M}(\rho_{k-1})
\: - \:
\op{tr}_\mathcal{M}(\xi_1)\otimes\cdots\otimes
\op{tr}_\mathcal{M}(\xi_{k-1})
\|_\mathrm{tr} < \delta(|x|)
\]
for some negligible $\delta(|x|)$ follows from the triangle inequality.

Finally, by applying the constructions from Lemmas~\ref{lemma:XOR2} and
\ref{lemma:amplify2} to $(Q_0,Q_1)$ appropriately results in circuits $R_0$
and $R_1$ that specify mixed states $\gamma_0$ and $\gamma_1$, respectively,
such that
\begin{enumerate}
\item[(i)]
$x\in A_\mathrm{yes} \Rightarrow \|\gamma_0 - \gamma_1\|_\mathrm{tr}<\alpha$,
and
\item[(ii)]
$x\in A_\mathrm{no} \Rightarrow \|\gamma_0 - \gamma_1\|_\mathrm{tr}> \beta$
\end{enumerate}
for any chosen constants $\alpha,\beta\in(0,1)$.

Thus, $x\in A_\mathrm{yes}$ implies
$(R_0,R_1)\in(\alpha,\beta)\mbox{-QSD}_\mathrm{no}$
and $x\in A_\mathrm{no}$ implies
$(R_0,R_1)\in(\alpha,\beta)\mbox{-QSD}_\mathrm{yes}$
as required.
\qed
\vspace{-2mm}

\begin{cor}
\label{cor:closed_under_complement}
{QSZK} is closed under complement.
\end{cor}

\begin{cor}
\label{cor:1-bit}
For any language $L \in \mbox{QSZK}$ there is a 2-message honest verifier
quantum statistical zero-knowledge proof system with exponentially
small completeness error and soundness error exponentially close to 1/2
in which the prover's message to the verifier consists of a single bit.
\end{cor}

\noindent
Corollary \ref{cor:closed_under_complement} follows from
Theorem~\ref{theorem:complete} together with
Theorem~\ref{theorem:complement_in_QSZK}, and 
Corollary~\ref{cor:1-bit} follows from Theorem~\ref{theorem:complete} and
the proof of Theorem~\ref{theorem:distance}.

\begin{cor}
\label{cor:QSZK_in_PSPACE}
{QSZK} $\subseteq$ {PSPACE}.
\end{cor}

\noindent
In order to prove this Corollary, let us consider the following problem.

\vspace{2mm}
\noindent
\underline{Trace Norm Approximation (TNA)}
\vspace{2mm}

\noindent
\begin{tabular}{@{}lp{5.5in}}
Input: & An $n\times n$ matrix $X$ (with entries having rational real and
imaginary parts) and an accuracy parameter $1^k$.\\[2mm]
Output: & A nonnegative rational number $r$ satisfying
$\left|\,r - \|X\|_\mathrm{tr}\,\right| \:<\: 2^{-k}$.
\end{tabular}

\begin{prop}
\label{prop:TNA_in_NC}
TNA $\in$ {NC}.
\end{prop}
{\bf Proof.} [Sketch]
Consider the following algorithm.\vspace{2mm}

\noindent
\begin{tabular}{@{}lp{6in}}
1. & Compute $Y = X X^{\dagger}$.\\[1mm]

2. & Compute the characteristic polynomial of $Y$ (the coefficients
will be real since $Y$ is necessarily Hermitian).\\[1mm]

3. & Calculate the $n$ roots $\lambda_1,\ldots,\lambda_n$ of the characteristic
polynomial of $Y$ to $O(k + \log n)$ bits of precision. \\[1mm]

4. & Compute $r = \frac{1}{2}\sum_{j=1}^n\sqrt{\lambda_j}$, where each square
root is approximated to $O(k + \log n)$ bits of precision, and output
$r$.\\[1mm]
\end{tabular}

\noindent
The output $r$ is an approximation to one-half the trace of
$\sqrt{X X^{\dagger}}$, which is $\|X\|_\mathrm{tr}$.
The approximation is correct to $O(k)$ bits of precision as required.
Each step can be performed in NC; simple arithmetic operations and
multiplication of matrices are well-known to be in NC, the fact that the
characteristic polynomial can be computed in NC was shown by
Csanky \cite{Csanky76}, and polynomial root approximation was shown to be in
NC by Neff \cite{Neff94}.
\qed

\noindent
{\bf Proof of Corollary~\ref{cor:QSZK_in_PSPACE}.}
[Sketch]
By Theorem~\ref{theorem:complete} it suffices to show that
$(\alpha,\beta)$-QSD is in {PSPACE}.
Recall that for any function $s(n)\geq \log n$, $\op{NC}(2^s)$ denotes the
class of languages computable by space $O(s)$-uniform boolean circuits having
size $2^{O(s)}$ and depth $s^{O(1)}$ \cite{BorodinC+83}.
The class $\op{NC}(2^s)$ is contained in $\op{DSPACE}(s^{O(1)})$
\cite{Borodin77}.
Thus, it will suffice to prove that $(\alpha,\beta)$-QSD is contained in
$\op{NC}(2^n)$.

Let $(Q_0,Q_1)$ be an input pair of quantum circuits specifying density
matrices $(\rho_0,\rho_1)$ on $k$ qubits, and let $n$ be the length of the
description of the pair $(Q_0,Q_1)$.
Obviously we may assume $k\leq n$, the number of qubits $m$ on which $Q_0$
and $Q_1$ act satisfies $m\leq n$, and each of $Q_0$ and $Q_1$ contains
at most $n$ gates.
We assume $Q_0$ and $Q_1$ are composed of gates that can be described by
unitary matrices having entries with rational real and imaginary parts (see
Section~\ref{sec:formalism}).
Thus, $\rho_0$ and $\rho_1$ correspond to $N\times N$ matrices where
$N\leq 2^n$, and for each entry of $\rho_0$ and $\rho_1$ the numerators and
denominators of the real and imaginary parts are $O(n)$-bit integers.

For each $i=0,1$ it is possible to compute $\ket{\psi_i} = Q_i|0^m\rangle$
(expressed as a $2^m$-dimensional vector with rational real and imaginary
parts) in $\op{NC}(2^n)$, simply by computing the product of the matrices
corresponding to each individual gate.
(In fact, there are better ways to do this from a complexity-theoretic
standpoint \cite{FortnowR99}, but this method is sufficient for our needs.)
Once these vectors are computed, it is possible to compute $\rho_0-\rho_1$ in
$\op{NC}(2^n)$ by constructing $\ket{\psi_0}\bra{\psi_0}$ and
$\ket{\psi_1}\bra{\psi_1}$, performing the partial trace on the non-output
qubits for each matrix (which involves computing a sum of at most $2^n$
matrices, each of which is obtained by multiplying $\ket{\psi_i}\bra{\psi_i}$
on the left and on the right by a $2^k\times2^m$ or $2^m\times2^k$ matrix,
respectively, as in the definition of the partial trace), and then computing
the difference of the resulting matrices.
Once we have $\rho_0 - \rho_1$, we may use the method described in
Proposition~\ref{prop:TNA_in_NC} to compute $\|\rho_0-\rho_1\|_\mathrm{tr}$
in $\op{NC}(2^n)$ (which is NC with respect to the size of $\rho_0 - \rho_1$).
Since it is only required that the cases
$\|\rho_0 - \rho_1\|_\mathrm{tr}\leq\alpha$ and
$\|\rho_0 - \rho_1\|_\mathrm{tr}\geq\beta$ be discriminated,
$\|\rho_0-\rho_1\|_\mathrm{tr}$ need in fact only be computed to $O(1)$ bits of
precision.
This completes the proof.
\qed
\vspace{-4mm}

%=============================================================================%

\section{Conclusion}
\label{sec:conclusion}

We have given a simple definition for honest verifier quantum statistical
zero-knowledge and proved several facts about the resulting complexity class.
Many questions regarding quantum statistical zero-knowledge, and quantum
zero-knowledge more generally, are left open.
For instance:
\begin{itemize}
\item
What are other natural definitions for quantum statistical zero-knowledge,
and how do they compare to our definition?
In particular, how does our definition for honest verifier quantum statistical
zero-knowledge compare to possible definitions for (not necessarily honest
verifier) quantum statistical zero-knowledge?
Are there quantum protocols that satisfy intuitive notions of
statistical zero-knowledge that do not satisfy our definition?

\item
What is the most reasonable definition for computational quantum
zero-knowledge, and what can be said about this class?

\item
What further relations among {QSZK} and other complexity classes can be shown?
Is there a better upper bound than {PSPACE}?
Is it possible that {NP} $\subseteq$ {QSZK}, or do unexpected consequences
result from such an assumption?

\item
The Quantum State Distinguishability problem is natural from the perspective
of quantum computation and quantum information theory, but is rather unnatural
outside of this scope.
Are there more natural problems that are candidates for problems in
{QSZK} but not in SZK?

\end{itemize}

\subsection*{Acknowledgments}

I thank Gilles Brassard and Claude Cr\'epeau for several discussions
concerning quantum zero-knowledge, and in particular for convincing me
of the difficulties in defining (not necessarily honest verifier) quantum
zero-knowledge.

%=============================================================================%

\bibliographystyle{plain}

%=============================================================================%

\newpage

\begin{center}
\Large\bf Appendix
\end{center}

\appendix

\section{Quantum circuits and the quantum formalism}

\label{sec:formalism}

We assume that the reader is familiar with the basics of quantum computation,
including the notion of (pure) quantum states, unitary operators, and
projective (or von Neumann) measurements.
We also assume familiarity with the quantum circuit model.
For further background information we refer the reader to
Nielsen and Chuang \cite{NielsenC00}, Berthiaume \cite{Berthiaume97},
and Kitaev \cite{Kitaev97}.
In this paper we will rely heavily on the so-called {\em density matrix
formalism}, which we briefly discuss below.
This formalism is discussed in detail by Nielsen and Chuang.

We use the following notion of a uniform family of quantum circuits.
A family $\{Q_x\}$ of quantum circuits is said to be {\em polynomial-time
uniformly generated} if there exists a deterministic procedure that, on
input $x$, outputs a description of $Q_x$ and runs in time polynomial in $x$.
It is assumed that the number of gates in any circuit is not more than the
length of that circuit's description (i.e., no compact descriptions of large
circuits are allowed), so that $Q_x$ must have size polynomial in $|x|$.
We also assume that quantum circuits are composed of gates from some
reasonable, universal, finite set of (unitary) gates.
By ``reasonable'' we mean, for instance, that gates cannot be defined
by matrices with non-computable, or difficult to compute, entries.
In fact, it will be helpful later to use the fact that any quantum
circuit composed of gates from any reasonable set of basis gates can
be efficiently simulated by a quantum circuit consisting only of gates
from a finite collection whose corresponding matrices have only entries
with rational real and imaginary parts.
See, for instance, Section 4.5.3 in Nielsen and Chuang for further
discussion.
It should be noted that our notion of uniformity is somewhat nonstandard,
since we allow an input $x$ to be given to the procedure generating the
circuits rather than just $|x|$ written in unary (with $x$ given
as input to the circuit itself).
This does not change the computational power for the resulting class of
quantum circuits, however, and we find that it is more convenient to
describe quantum interactive proof systems using this notion.

Now we briefly discuss the density matrix formalism.
Among other things, this formalism provides a way to describe subsystems of
quantum systems, which is helpful when considering quantum interactive proof
systems and crucial for extending the notion of zero-knowledge to the quantum
setting.

Recall that a {\em pure (quantum) state} (or superposition) of an $n$-qubit
quantum system is a unit vector in the Hilbert space\footnote{All Hilbert
spaces referred to in this paper are assumed to be finite dimensional.}
$\mathcal{H}=\ell_2(\{0,1\}^n)$, and corresponding to each pure state
$\ket{\psi}\in\mathcal{H}$ is a
linear functional $\bra{\psi}$ that maps each vector $\ket{\phi}$ to the inner
product $\bracket{\psi}{\phi}$ (conjugate-linear in the first argument).
A {\em mixed state} of a quantum system is a state that may be
described by a distribution on (not necessarily orthogonal) pure states.
A collection $\{\left(p_k,\ket{\psi_k}\right)\}$ such that $0\leq p_k$,
$\sum_k p_k=1$, and each $\ket{\psi_k}$ is a pure state is called a
{\em mixture}: for each $k$, the system is in state $\ket{\psi_k}$
with probability $p_k$.
For a given mixture $\{\left(p_k,\ket{\psi_k}\right)\}$, we associate a
{\em density matrix} $\rho$ having operator representation
$\rho = \sum_k p_k\ket{\psi_k}\bra{\psi_k}$.
Necessary and sufficient conditions for a given matrix $\rho$ to be a density
matrix (i.e., to represent some mixed state) are (i)~$\rho$ must be positive
semidefinite, and (ii)~$\rho$ must have unit trace.
Two mixtures can be distinguished (in a statistical sense) if and only if
they yield different density matrices, and for this reason we interpret a
given density matrix $\rho$ as being a canonical representation of a given
mixed state.
Unitary transformations and measurements work as follows on density matrices.
Applying a unitary operator $U$ to $\rho$ yields $U\rho U^{\dagger}$, and
measuring a mixed state $\rho$ according to a (projective) measurement
described by some complete, orthogonal set of projections
$\{\Pi_1,\ldots,\Pi_l\}$ yields result $j$ with probability $\op{tr}\Pi_j\rho$.

The quantum circuit model has been extended to the density matrix formalism by
Aharonov, Kitaev, and Nisan~\cite{AharonovK+98}, who show that the
the resulting model (which allows more general types of gates than the usual
model, such as ``measurement gates'') is equivalent in power to the usual
model in which only unitary gates are allowed.
As stated above, we assume all quantum circuits in our model consist of
only unitary gates, which causes no loss of generality following from
this equivalence.

In order to describe the density matrix formalism further, it will be
helpful at this point to introduce some notation.
For a given Hilbert space $\mathcal{H}$, let $\mathbf{L}(\mathcal{H})$ denote
the set of linear operators on $\mathcal{H}$, let $\mathbf{D}(\mathcal{H})$
denote the set of positive semidefinite operators on $\mathcal{H}$ having unit
trace (so that $\mathbf{D}(\mathcal{H})$ may be identified with the set of
mixed states of a given system), let $\mathbf{U}(\mathcal{H})$ denote the set
of unitary operators on $\mathcal{H}$, and let $\mathbf{P}(\mathcal{H})$ denote
the set of projection operators on $\mathcal{H}$.

Given Hilbert spaces $\mathcal{H}$ and $\mathcal{K}$, we define a mapping
$\op{tr}_\mathcal{K}:\mathbf{D}(\mathcal{H}\otimes\mathcal{K})
\rightarrow\mathbf{D}(\mathcal{H})$ as follows:
\[
\op{tr}_{\mathcal{K}}\rho =
\sum_{j=1}^n (I\otimes\bra{e_j})\rho(I\otimes\ket{e_j}),
\]
where $\{\ket{e_1},\ldots,\ket{e_n}\}$ is any orthonormal basis of
$\mathcal{K}$.
This mapping is known as the {\em partial trace}, and has the following
intuitive meaning: given a mixed state
$\rho\in\mathbf{D}(\mathcal{H}\otimes\mathcal{K})$ of a bipartite system
(meaning that the first part of the system corresponds to $\mathcal{H}$ and
the second part to $\mathcal{K}$), $\op{tr}_{\mathcal{K}}\rho$ is the mixed
state of the first part of the system obtained by discarding or not
considering the second part of the system.
To say that a particular part of a quantum system is {\em traced out} means
that the partial trace is performed, removing this part of the system from
consideration.

A {\em purification} of a given mixed state $\rho\in\mathbf{D}(\mathcal{H})$
is any pure state $\ket{\psi}$ of a larger quantum system that gives $\rho$
when part of the system is traced out.
In other words, we have $\ket{\psi}\in\mathcal{H}\otimes\mathcal{K}$ for some
Hilbert space $\mathcal{K}$ such that
$\op{tr}_\mathcal{K}\ket{\psi}\bra{\psi} = \rho$.

For $X\in\mathbf{L}(\mathcal{H})$ define
\[
\|X\|_{\mathrm{tr}}\:=\:\frac{1}{2}\op{tr}\sqrt{X^{\dagger}X}.
\]
(Recall that for any positive semidefinite matrix $A$ there is a unique
positive semidefinite matrix denoted $\sqrt{A}$ that satisfies
$(\sqrt{A})^2 = A$.)
The function $\|\cdot\|_\mathrm{tr}$ is a norm called the {\em trace norm},
and generalizes the norm induced by the statistical difference or total
variation distance (i.e., one-half the $\ell_1$ norm).
For any normal matrix $X$, the trace norm is simply one-half the sum of the
absolute values of the eigenvalues of $X$.
For any $X\in\mathbf{L}(\mathcal{H})$ we have
$\|X\|_\mathrm{tr} = \max_A|\op{tr}AX|$,
where the maximum is over all positive semidefinite
$A\in\mathbf{L}(\mathcal{H})$ with $\|A\|\leq 1$.
Alternately we may take the maximum to be over all projections
$A\in\mathbf{P}(\mathcal{H})$, which does not change the maximum value.

Given two mixed states $\rho,\xi\in\mathbf{D}(\mathcal{H})$, define
the {\em fidelity} of $\rho$ and $\xi$ by
\[
F(\rho,\xi) = \op{tr}\sqrt{\rho^{1/2}\,\xi\,\rho^{1/2}}.
\]
For all $\rho,\xi\in\mathbf{D}(\mathcal{H})$ we have
$1-F(\rho,\xi)\leq\|\rho-\xi\|_\mathrm{tr}\leq\sqrt{1-F(\rho,\xi)^2}$.
This and several other facts about the trace norm and the fidelity are
discussed in the next section.

%=============================================================================%

\section{Basic properties of fidelity and the trace distance}

In this section of the appendix we give proofs or references for basic facts
about trace distance and fidelity that are used elsewhere in the paper.

\begin{prop}
\label{prop:fidelity}
For all $\rho,\xi\in\mathbf{D}(\mathcal{H})$ we have
\[
1 - F(\rho,\xi) \leq \|\rho - \xi\|_\mathrm{tr} \leq \sqrt{1 - F(\rho,\xi)^2}.
\]
\end{prop}

\noindent
See Section 9.2.3 of Nielsen and Chuang \cite{NielsenC00} for a proof.

\begin{prop}{\label{prop:fidelity_mult}}
For any $\rho_1,\xi_1\in\mathbf{D}(\mathcal{H})$ and
$\rho_2,\xi_2\in\mathbf{D}(\mathcal{K})$ we have
\[
F(\rho_1\otimes\rho_2,\xi_1\otimes\xi_2) = F(\rho_1,\xi_1)\,F(\rho_2,\xi_2).
\]
\end{prop}
{\bf Proof.}
For given positive semidefinite matrices $A$ and $B$ we have
$\sqrt{A\otimes B} = \sqrt{A}\otimes\sqrt{B}$ and
$\op{tr}A\otimes B = (\op{tr}A)(\op{tr}B)$.
Thus,
\begin{eqnarray*}
F(\rho_1\otimes\rho_2,\xi_1\otimes\xi_2)
& = & \op{tr}\sqrt{(\rho_1\otimes\rho_2)^{1/2}\,(\xi_1\otimes\xi_2)\,
(\rho_1\otimes\rho_2)^{1/2}}\\[2mm]
& = & \op{tr}\left(\sqrt{\rho_1^{1/2}\,\xi_1\,\rho_1^{1/2}}
\otimes \sqrt{\rho_2^{1/2}\,\xi_2\,\rho_2^{1/2}}\right)\\[2mm]
& = & F(\rho_1,\xi_1)\,F(\rho_2,\xi_2)
\end{eqnarray*}
as required.
\qed

\begin{prop}{\label{prop:trace_mult}}
Let $A\in\mathrm{L}(\mathcal{H})$ and $B\in\mathrm{L}(\mathcal{K})$.
Then $\|A\otimes B\|_{\mathrm{tr}}=2\,\|A\|_{\mathrm{tr}}\|B\|_{\mathrm{tr}}$.
\end{prop}
\noindent {\bf Proof.}
We have
\begin{eqnarray*}
\|A\otimes B\|_{\mathrm{tr}} & = & \frac{1}{2}\op{tr}\sqrt{A^{\dagger}A
\otimes B^{\dagger}B}\\[1mm]
& = & \frac{1}{2}\op{tr}\sqrt{A^{\dagger}A}\otimes \sqrt{B^{\dagger}B}\\
& = & \frac{1}{2}\left(\op{tr}\sqrt{A^{\dagger}A}\right)
\left(\op{tr}\sqrt{B^{\dagger}B}\right)\\
& = & 2\,\|A\|_{\mathrm{tr}} \|B\|_{\mathrm{tr}}
\end{eqnarray*}
as required.
\qed

\begin{prop}{\label{prop:XOR1}}
Let $\rho_0,\rho_1\in\mathbf{D}(\mathcal{H})$ and
$\xi_0,\xi_1\in\mathbf{D}(\mathcal{K})$.
Define
\begin{eqnarray*}
\gamma_0 & = & \frac{1}{2}(\rho_0\otimes\xi_0)+\frac{1}{2}(\rho_1\otimes
\xi_1),\\[2mm]
\gamma_1 & = & \frac{1}{2}(\rho_0\otimes\xi_1)+\frac{1}{2}(\rho_1\otimes
\xi_0).
\end{eqnarray*}
Then $\|\gamma_0 - \gamma_1\|_{\mathrm{tr}} =
\|\rho_0 - \rho_1\|_{\mathrm{tr}}\,\|\xi_0 - \xi_1\|_{\mathrm{tr}}$.
\end{prop}
\noindent
{\bf Proof.}
We have
\begin{eqnarray*}
\|\gamma_0 - \gamma_1\|_{\mathrm{tr}} & = &
\left\| \frac{1}{2}(\rho_0\otimes\xi_0)+\frac{1}{2}(\rho_1\otimes\xi_1)
- \frac{1}{2}(\rho_0\otimes\xi_1)-\frac{1}{2}(\rho_1\otimes\xi_0)
\right\|_{\mathrm{tr}}\\
& = & \left\|\frac{1}{2}(\rho_0 - \rho_1)\otimes(\xi_0 - \xi_1)
\right\|_{\mathrm{tr}}\\
& = &
\|\rho_0 - \rho_1\|_{\mathrm{tr}}\cdot\|\xi_0 - \xi_1\|_{\mathrm{tr}}
\end{eqnarray*}
as required.
\qed

\begin{prop}{\label{prop:tensor_distance}}
Let $\rho_0,\rho_1\in\mathbf{D}(\mathcal{H})$ and
$\xi_0,\xi_1\in\mathbf{D}(\mathcal{K})$.
Then
\[
\|\rho_0\otimes\xi_0 - \rho_1\otimes\xi_1\|_\mathrm{tr}
\leq
\|\rho_0 - \rho_1\|_\mathrm{tr}
+
\|\xi_0 - \xi_1\|_\mathrm{tr}.
\]
\end{prop}
\noindent
{\bf Proof.}
We have
\begin{eqnarray*}
\|\rho_0\otimes\xi_0 - \rho_1\otimes\xi_1\|_\mathrm{tr}
& \leq &
\|\rho_0\otimes\xi_0 - \rho_1\otimes\xi_0\|_\mathrm{tr}
+
\|\rho_1\otimes\xi_0 - \rho_1\otimes\xi_1\|_\mathrm{tr}\\
& = &
\|(\rho_0 - \rho_1)\otimes\xi_0\|_\mathrm{tr}
+
\|\rho_1\otimes(\xi_0 - \xi_1)\|_\mathrm{tr}\\
& = & 
\|\rho_0 - \rho_1\|_\mathrm{tr} + \|\xi_0 - \xi_1\|_\mathrm{tr}.
\end{eqnarray*}
as required.
\qed

\begin{theorem}{\label{theorem:Schmidt}}
Let $\ket{\phi},\ket{\psi}\in\mathcal{H}\otimes\mathcal{K}$ satisfy
$\op{tr}_\mathcal{K}\ket{\phi}\bra{\phi} =
\op{tr}_\mathcal{K}\ket{\psi}\bra{\psi} = \rho$
for some $\rho\in\mathbf{D}(\mathcal{H})$.
Then there exists $U\in\mathbf{U}(\mathcal{K})$ such that
$(I\otimes U)\ket{\phi} = \ket{\psi}$.
\end{theorem}

\noindent
See Section 2.5 of Nielsen and Chuang \cite{NielsenC00} for a proof.

\begin{theorem}
\label{theorem:Uhlmann}
Let $\rho,\xi\in\mathbf{D}(\mathcal{H})$, and let $\mathcal{K}$ be such that
there exist purifications
$\ket{\phi_0},\ket{\psi_0}\in\mathcal{H}\otimes\mathcal{K}$ of $\rho$ and
$\xi$, respectively (i.e.,
$\op{tr}_{\mathcal{K}}\ket{\phi_0}\bra{\phi_0} = \rho$ and
$\op{tr}_{\mathcal{K}}\ket{\psi_0}\bra{\psi_0} = \xi$).
Then
\[
F(\rho,\xi) = \max_{\ket{\phi},\ket{\psi}}|\langle \phi|\psi\rangle|,
\]
where the maximum is over all purifications
$\ket{\phi},\ket{\psi}\in\mathcal{H}\otimes\mathcal{K}$ of $\rho$ and $\xi$,
respectively.
\end{theorem}

\noindent
See Section 9.2.2 of Nielsen and Chuang \cite{NielsenC00} for a proof.

\begin{lemma}
\label{cor:of_Uhlmann}
Let $\rho,\xi\in\mathbf{D}(\mathcal{H})$ and let
$\sigma\in\mathbf{D}(\mathcal{H}\otimes\mathcal{K})$ satisfy
$\op{tr}_\mathcal{K}\sigma = \rho$.
Let $\ket{\psi}\in\mathcal{H}\otimes\mathcal{K}$ be a purification of
$\xi$, i.e., $\op{tr}_{\mathcal{K}}\ket{\psi}\bra{\psi} = \xi$.
Then $\bra{\psi}\rho\ket{\psi}\leq F(\rho,\xi)^2$.
\end{lemma}

\noindent
{\bf Proof.}
We have
\[
\sqrt{\langle\psi|\rho|\psi\rangle} \: = \: F(\ket{\psi}\bra{\psi},\rho)
\: = \: \max_{\ket{\phi_0},\ket{\psi_0}}|\langle\phi_0|\psi_0\rangle|
\:\leq\: F(\rho,\xi).
\]
Here the maximum is over purifications of $\rho$ and $\ket{\psi}\bra{\psi}$.
The inequality follows from the fact that any purification of
$\ket{\psi}\bra{\psi}$ is also a purification of $\xi$.
\qed

\begin{lemma}{\label{lemma:amplify}}
Let $\rho,\,\xi\in\mathbf{D}(\mathcal{H})$ satisfy
$\|\rho - \xi\|_\mathrm{tr} = \varepsilon$.
Then
\[
1 - e^{-k\varepsilon^2/2}\:<\:\|\rho^{\otimes k}-\xi^{\otimes k}\|_\mathrm{tr}
\:\leq\: k\varepsilon.
\]
\end{lemma}
{\bf Proof.}
The second inequality follows immediately from
Proposition~\ref{prop:tensor_distance}.
Let us prove the first inequality.
We have
\begin{multline*}
\|\rho^{\otimes k}-\xi^{\otimes k}\|_\mathrm{tr}
\geq 1-F(\rho^{\otimes k},\xi^{\otimes k})
= 1-F(\rho,\xi)^k
\geq 1-\left(\sqrt{1 - \|\rho - \xi\|_\mathrm{tr}^2}\right)^k\\
= 1-(1-\varepsilon^2)^\frac{k}{2}
= 1-(1-\varepsilon^2)^{\frac{1}{\varepsilon^2}
\cdot\frac{k\varepsilon^2}{2}}
> 1 - e^{-\frac{k\varepsilon^2}{2}}
\end{multline*}
as required.
\qed

\begin{lemma}{\label{lemma:closeness1}}
Let $\rho,\xi\in\mathbf{D}(\mathcal{H})$ satisfy
$F(\rho,\xi)\geq 1 - \varepsilon$ and let
$\ket{\phi},\ket{\psi}\in\mathcal{H}\otimes\mathcal{K}$ be purifications
of $\rho$ and $\xi$, respectively, i.e.,
$\op{tr}_{\mathcal{K}}\ket{\phi}\bra{\phi} = \rho$ and
$\op{tr}_{\mathcal{K}}\ket{\psi}\bra{\psi} = \xi$.
Then there exists $U\in\mathbf{U}(\mathcal{K})$ such that
\[
\| (I\otimes U) \ket{\phi} - \ket{\psi} \| \leq \sqrt{2\varepsilon}.
\]
\end{lemma}
\noindent
{\bf Proof.}
By Theorem~\ref{theorem:Uhlmann} we have
\[
F(\rho,\xi)=\max_{\ket{\phi_0},\ket{\psi_0}}|\langle\phi_0|\psi_0\rangle|,
\]
where the maximum is over all purifications
$\ket{\phi_0},\ket{\psi_0}\in\mathcal{K}$ of $\rho$ and $\xi$, respectively.
Let $\ket{\phi_0}$ and $\ket{\psi_0}$ be pure states achieving this maximum,
and assume without loss of generality that $\langle\phi_0|\psi_0\rangle$
is a nonnegative real number.

Since $\ket{\phi}$ and $\ket{\phi_0}$ are both purifications of
$\rho$, we have by Theorem~\ref{theorem:Schmidt} that there exists
\mbox{$V\in\mathbf{U}(\mathcal{K})$} such that
$\ket{\phi_0}=(I\otimes V)\ket{\phi}$.
Similarly, there exists
$W\in\mathbf{U}(\mathcal{K})$ such that
\mbox{$\ket{\psi_0}=(I\otimes W)\ket{\psi}$}.

Define $U = V^{\dagger}W$.
Then
\[
\|(I\otimes U)\ket{\phi} - \ket{\psi}\|
= \|(I\otimes W)\ket{\phi} - (I\otimes V)\ket{\psi}\|
= \|\ket{\phi_0} - \ket{\psi_0}\|
= \sqrt{2 - 2\langle \phi_0 |\psi_0\rangle}
\leq \sqrt{2\varepsilon}
\]
as required.
\qed

%=============================================================================%

\end{document}